\documentclass[journal,10pt]{IEEEtran}
\IEEEoverridecommandlockouts
\usepackage[utf8]{inputenc}

\usepackage{tikz,pgfplots}
\usetikzlibrary{positioning, calc,backgrounds}
\usepgfplotslibrary{fillbetween} 
\usetikzlibrary{patterns} 
\pgfplotsset{compat=1.18}
\usepackage{pgfplotstable}
\usepackage{ifthen} 
\usepackage{xcolor}
\definecolor{mygreen}{HTML}{A1A9D0} 
\definecolor{myyellow}{HTML}{F0988C} 
\definecolor{myorange}{HTML}{9E9E9E} 
\definecolor{myred}{HTML}{B883D4}
\usepackage{tcolorbox}
\usetikzlibrary{arrows.meta, fit, backgrounds, decorations.pathreplacing}
\tcbuselibrary{most}
\usetikzlibrary{shadows}
\usetikzlibrary{shapes}
\usetikzlibrary{shapes.arrows}
\usetikzlibrary{shapes.geometric}
\usepgfplotslibrary{groupplots}

\usepackage{enumitem}
\usepackage{graphicx} % Required for inserting images
\usepackage{bm}
\usepackage{amsthm}
\usepackage{amsmath}
\usepackage{amsfonts}
\usepackage{algorithm2e}
\usepackage{algpseudocode}
\usepackage{multirow}
\usepackage{amssymb}
\usepackage{pifont}
\usepackage{subfigure}
\usepackage{soul}
\usepackage{tabularx}
\newcolumntype{Y}{>{\centering\arraybackslash}X}
\newcolumntype{K}[1]{>{\centering\arraybackslash}m{#1}}
\newcolumntype{L}[1]{>{\centering\arraybackslash}p{#1}}
\usepackage{multirow}
\usepackage{xcolor}
\usepackage{marginnote}
\usepackage{cite}
\usepackage{bm}
\usepackage{mathtools}
\usepackage{mathrsfs}
\usepackage{euscript}
\usepackage[hidelinks]{hyperref}

\author{Zihang~Song,~\IEEEmembership{Member,~IEEE,}
        Matteo~Zecchin,~\IEEEmembership{Member,~IEEE,}
        Bipin~Rajendran,~\IEEEmembership{Senior Member,~IEEE,}
        Petar~Popovski,~\IEEEmembership{Fellow,~IEEE,}
        and~Osvaldo~Simeone,~\IEEEmembership{Fellow,~IEEE}
\thanks{
	Z. Song (zsong@es.aau.dk) and P. Popovski (petarp@es.aau.dk) are with the Connectivity Section, Department of Electronic Systems, Aalborg University, 9220 Aalborg, Denmark.
    M. Zecchin (zecchin@eurecom.fr) is with the Communication Systems Department, EURECOM, 06410 Sophia Antipolis, France.
    B. Rajendran (b.rajendran@nulondon.ac.uk) and O. Simeone (o.simeone@nulondon.ac.uk) are with the Institute for Intelligent Networked Systems (INSI), Northeastern University London, 1 Portsoken Street, London E1 8PH, United Kingdom.
	
	The work was supported by the European Union's Horizon Europe project CENTRIC (101096379), the European Research Council (ERC) under grant No. 101198347, the EPSRC project (EP/X011852/1), the Open Fellowships of the EPSRC (EP/W024101/1 and EP/X011356/1), the Marie Skłodowska-Curie Action (MSCA) Postdoc Fellowships project NEUER (No. 101206861), and the Velux Foundation, Denmark, through the Villum Investigator Grant WATER (37793). Views and opinions expressed are however those of the author(s) only and do not necessarily reflect those of the European Union or the European Research Council. Neither the European Union nor the granting authority can be held responsible for them. }
}

\begin{document}
\title{In-Context Learning for MIMO Turbo Receivers}
\maketitle

\begin{abstract}
Multiple-input multiple-output (MIMO) turbo receivers iteratively exchange information between a soft-input soft-output detector and a decoder. Traditional implementations rely on pilot-based channel estimates, which can be poor in the case of short pilot sequences or mismatched channel models, resulting in significant performance degradation. To address this limitation, we introduce the EXTrinsic In-Context Learning Network (EXT-ICLNet), a Transformer-based detector that takes the pilots as in-context examples and is trained to approximate the symbol extrinsic posterior distribution given the pilots, without explicit channel estimation. EXT-ICLNet is thus compatible with existing turbo receivers, as it implements a log-domain prior skip connection, ensuring that the network learns only the extrinsic information. Numerical results for a $4\times4$ MIMO system with 16-QAM and a quantized front end show that EXT-ICLNet outperforms baselines based on explicit channel state information (CSI) estimation at low-to-medium signal-to-noise ratios (SNRs), as well as expectation-propagation (EP) baselines with perfect CSI at high SNRs. The results are further validated by comparing the extrinsic information transfer (EXIT) charts of EXT-ICLNet, a neural EP baseline, and EXT-ICLNet without skip connection, demonstrating how EXT-ICLNet provides a more favorable transfer of a priori and extrinsic information.
\end{abstract}

\section{Introduction}

\subsection{Context and Motivation}
 
Turbo receivers are a class of iterative receivers that refine their estimates of the transmitted bits by repeated exchange of information between a soft-input soft-output detector and a decoder. These receivers have been shown to approach the performance of optimal joint detection and decoding in coded systems~\cite{koetter2004turbo,douillard1995iterative,tuchler2011turbo,chen2024memory,direnna2024iterative,chi2026achievable}.

\begin{figure}[t]
    \centering
    % system_model_1.tex -- single codeword, interleaved, split across N_t streams
% Requires tikz libraries: arrows.meta, calc, positioning
\begin{tikzpicture}[
    >={Latex[length=1.6mm, width=1.4mm]},
    font=\footnotesize,
    line width=0.5pt,
    block/.style={draw, align=center, fill=white, inner sep=4pt,thick,
                  rounded corners=2pt, minimum height=0.6cm},
    smallblock/.style={draw, align=center, fill=white, inner sep=4pt,thick,
                       rounded corners=2pt, minimum height=0.5cm},
    arrow/.style={->, thick},
    edgelbl/.style={inner sep=1.5pt, font=\scriptsize},
    sumcircle/.style={
        draw, circle, minimum size=0.32cm, inner sep=0pt, fill=white,
        font=\scriptsize,thick
    }
]

% === ADC (B-bit quantizer) pic ===
\tikzset{
    adc/.pic={
        \draw[line width=0.5pt,thick] (0,0) -- (0.1,0);
        \draw[line width=0.5pt,thick] (0.1,0) -- (0.45,0.3);
        \draw[line width=0.5pt,thick] (0.55,0) -- (0.65,0);
        \draw[->, line width=0.4pt,thick] (0.1,0.3) .. controls (0.25,0.18) .. (0.32,0);
        \node[font=\scriptsize] at (0.33,-0.15) {$B$-bit};
    }
}

% === Encoder, interleaver, serial-to-parallel ===
\node[smallblock, inner sep=2pt] (encoder) {Enc.};
\draw[arrow] ($(encoder.west)+(-0.35,0)$)
    -- node[above, edgelbl] {$\bm{a}$} (encoder.west);

\node[smallblock, right=0.25cm of encoder] (interleaver) {$\Pi$};
\draw[arrow] (encoder.east) -- (interleaver.west);

\node[block, minimum height=1.9cm, inner sep=2pt, right=0.35cm of interleaver] (sp) {S/P};
\draw[arrow] (interleaver.east) -- node[above, edgelbl] {$\bm{c}$} (sp.west);

% === QAM mappers ===
\node[smallblock, minimum width=1.1cm, inner sep=2pt, anchor=west]
    (qam1) at ($(sp.east)+(0.3,0.7)$) {$M$-QAM};
\node[smallblock, minimum width=1.1cm, inner sep=2pt, anchor=west]
    (qam2) at ($(sp.east)+(0.3,-0.7)$) {$M$-QAM};
\node at ($(qam1)!0.5!(qam2)$) {$\vdots$};

\draw[arrow] (sp.east|-qam1) -- (qam1.west);
\draw[arrow] (sp.east|-qam2) -- (qam2.west);

% === Channel ===
\node[block, minimum height=1.9cm, anchor=west]
    (channel) at ($(qam1.east|-sp)+(0.55,0)$) {Channel\\$\bm{H}$};

\draw[arrow] (qam1.east) -- node[above=0.2cm, edgelbl] {$x_i^{(1)}$}
    (qam1.east-|channel.west);
\draw[arrow] (qam2.east) -- node[above=0.2cm, edgelbl] {$x_i^{(N_t)}$}
    (qam2.east-|channel.west);

% === Channel output ports + noise summers + ADCs ===
\coordinate (chout1) at ($(channel.east)+(0,0.7)$);
\coordinate (chout2) at ($(channel.east)+(0,-0.7)$);

\node[sumcircle, anchor=west] (sum1) at ($(chout1)+(0.45,0)$) {$+$};
\node[sumcircle, anchor=west] (sum2) at ($(chout2)+(0.45,0)$) {$+$};

\node[anchor=west] (dots4) at ($(channel.east)+(0.95,0.2)$) {$\vdots$};

\draw[arrow] (chout1) -- (sum1.west);
\draw[arrow] (chout2) -- (sum2.west);

\draw[arrow] ($(sum1.north)+(0,0.45)$) node[above, edgelbl] {$n_i^{(1)}$}
    -- (sum1.north);
\draw[arrow] ($(sum2.north)+(0,0.45)$) node[above, edgelbl] {$n_i^{(N_r)}$}
    -- (sum2.north);

\pic[anchor=west,thick] (adc1) at ($(sum1.east)+(0.15,0)$) {adc};
\pic[anchor=west,thick] (adc2) at ($(sum2.east)+(0.15,0)$) {adc};
\draw[line width=0.5pt,thick] (sum1.east) -- ($(sum1.east)+(0.15,0)$);
\draw[line width=0.5pt,thick] (sum2.east) -- ($(sum2.east)+(0.15,0)$);

\draw[arrow] ($(sum1.east)+(0.80,0)$)
    -- node[above, edgelbl] {$y_i^{(1)}$} ($(sum1.east)+(1.20,0)$);
\draw[arrow] ($(sum2.east)+(0.80,0)$)
    -- node[above, edgelbl] {$y_i^{(N_r)}$} ($(sum2.east)+(1.20,0)$);

\end{tikzpicture}
    \caption{Block diagram of coded MIMO transmission with a quantized front end.}
    \label{fig:system_model_1}\vspace{-3mm}
\end{figure}

In multiple-input multiple-output (MIMO) systems, the role of the detector is to equalize the received signal and return extrinsic information, i.e., information gained beyond the a priori information, which takes the form of bit-wise log-likelihood ratios (LLRs). Conventional detectors compute the extrinsic LLRs based on channel state information (CSI) obtained from pilot signals~\cite{tuchler2011turbo}. The CSI is typically a point estimate of the channel, in some receivers complemented by a Gaussian model of the estimation error~\cite{krebs2015error,jiang2015soft}. However, when pilot sequences are short or the receiver front end is coarsely quantized~\cite{choi2016near,mo2015capacity,liu2019low}, such estimates can be inaccurate, and the detection performance degrades substantially.

Neural detectors provide a data-driven alternative to model-based detectors, allowing the mapping from the received signal to symbol estimates to be learned directly from data~\cite{zhou2023graph,nguyen2023leveraging,hu2023understanding,he2024message,zhang2025cross}. However, their use in turbo receivers presents two main challenges. First, existing neural detectors~\cite{samuel2019learning,he2020model,khani2020adaptive,goutay2020deep,wei2020learned,pratik2021remimo,zhang2021meta,kosasih2022graph,zhou2023graph,hu2023understanding,sun2023robust,he2024message} rely on a point estimate of the CSI and therefore remain susceptible to channel-estimation errors. Second, turbo receivers require the detector to incorporate the a priori information provided by the decoder while returning only extrinsic information, i.e., excluding the corresponding a priori contribution. Standard neural detectors do not inherently enforce this separation, since the received signal and a priori information are jointly processed to produce posterior estimates \cite{koike2020neural,zhou2023graph}. 

To address the outlined limitations of existing approaches, we introduce the EXTrinsic In-Context Learning Network (EXT-ICLNet), a soft-input soft-output neural detector for MIMO turbo receivers (see Fig.~\ref{fig:system_model_1}) that leverages the in-context learning (ICL) capability of sequence models~\cite{brown2020language,garg2022can,akyurek2023what,vonoswald2023transformers,bai2023transformers} to avoid explicit CSI estimation. EXT-ICLNet uses the pilot sequence as in-context examples~\cite{zecchin2023context,rajagopalan2023transformers,zecchin2024cell,song2024neuromorphic,song2024transformer,fan2024decision,song2025context,zecchin2026context}, and directly processes the received signal together with the a priori information to approximate the symbol posterior distribution~\cite{muller2022transformers} (see Fig.~\ref{fig:turbo_receivers}). To ensure that the extrinsic output does not contain the corresponding a priori contribution, EXT-ICLNet incorporates a log-domain prior skip connection, as illustrated in Fig.~\ref{fig:architecture}. This architecture enables EXT-ICLNet to be trained end-to-end in a single stage with a cross-entropy loss on the transmitted symbols, unlike the specialized training procedures required by~\cite{koike2020neural,huang2021extrinsic,wiesmayr2022duidd,zhou2023graph,clausius2023component,clausius2025joint}.

\subsection{Related Work}\label{subse:related_work}

\emph{Model-Based MIMO Turbo Receivers}: As reviewed in~\cite{tuchler2011turbo}, representative soft-input soft-output detectors for MIMO include linear minimum mean square error detection with parallel interference cancellation (LMMSE-PIC)~\cite{tuchler2002minimum,studer2011asic}, the single tree-search sphere decoder (STS-SD)~\cite{hochwald2003achieving,studer2006soft,studer2010soft}, low-rank methods~\cite{dietl2007complexity}, Kalman-based frameworks~\cite{park2015iterative}, and expectation-propagation (EP)-based detectors~\cite{santos2018turbo,santos2019self}. Extensions to low-resolution analog-to-digital converter (ADC) systems use the Bussgang decomposition~\cite{liu2019low,li2017channel}. These detectors operate on a channel estimate obtained from the pilots. MIMO turbo receivers typically refine this estimate across turbo iterations by decision-directed recursive least squares (RLS) updates~\cite{da2021adaptive,shiao2008combined}, and some add the covariance of the channel-estimation error to the noise covariance of the linear-Gaussian residual model~\cite{krebs2015error,jiang2015soft}, so the estimation error enters detection only through a Gaussian approximation. Robust minimax turbo equalizers~\cite{kalantarova2011robust} and joint channel-estimation and decoding methods~\cite{sun2018joint,sun2019joint} account for the channel uncertainty in the detector design but are developed for single-stream systems.
 
\emph{Neural Detectors}: Deep-unfolded detectors map the iterations of a detection algorithm onto network layers with learnable parameters. OAMP-Net2, an unfolded orthogonal approximate message passing (OAMP) detector~\cite{he2020model}, and DetNet, an unfolded projected gradient detector~\cite{samuel2019learning}, take a channel estimate as input, and the deep expectation-maximization detector of~\cite{zhang2022deep} estimates the channel jointly with the data. Meta-learned~\cite{park2020learning} and adaptive~\cite{raviv2024adaptive,raviv2024modular} neural receivers adapt to the channel by updating the network weights from pilots, whereas EXT-ICLNet adapts to each channel through the pilot context with fixed weights.

\emph{Extrinsic Information in Neural Turbo Receivers}: EXT-GEPNet~\cite{zhou2023graph}, an EP detector enhanced with a graph neural network (GNN), obtains extrinsic outputs through a three-step procedure that trains an a posteriori probability network, generates extrinsic labels with masked priors, and retrains the network on these labels. Other approaches use nested residual architectures with min-max losses~\cite{koike2020neural}, open-loop training~\cite{huang2021extrinsic}, and learned iteration-dependent LLR interpolation~\cite{wiesmayr2022duidd}. These methods rely on specialized training or inference procedures to obtain extrinsic outputs from networks that process a priori information.

\emph{ICL-Based Symbol Detectors}: In ICL, a pretrained sequence model adapts to a new task by reading a prompt consisting of a context (a set of labeled examples) and a query, without updating its weights~\cite{garg2022can}. This capability was formalized as amortized Bayesian inference in~\cite{muller2022transformers}, where a Transformer trained with cross-entropy on data sampled from a task distribution recovers the posterior predictive distribution at the population optimum. In ICL-based MIMO detection, the task is the unknown channel, the context consists of the pilots, and the query is the received signal, so the detector operates without a point channel estimate~\cite{zecchin2023context,rajagopalan2023transformers,zecchin2026context}. ICL-based detection has since been extended to cell-free MIMO~\cite{zecchin2024cell}, frequency-selective channels~\cite{song2024transformer}, and decision-feedback schemes~\cite{fan2024decision}, and realized with neuromorphic computing~\cite{song2024neuromorphic,song2025xpikeformer} and state-space architectures~\cite{song2025context}. However, these detectors perform one-shot detection and do not exchange extrinsic information with a decoder.

\subsection{Main Contributions}

\begin{figure}[t]
    \centering
    \subfigure[]{
        % turbo_receiver_conv.tex -- restyled to match the architecture panels
% (same fonts, Latex arrowheads, rounded corners, and white fills).
% Requires tikz libraries: arrows.meta, positioning, calc
\begin{tikzpicture}[
    >={Latex[length=1.8mm]},
    font=\fontsize{8pt}{9pt}\selectfont,
    basenode/.style={draw, rounded corners=2pt, thick, align=center,
                     fill=white, minimum height=0.8cm},
    smallnode/.style={draw, rounded corners=2pt, thick, align=center,
                      fill=white, minimum width=0.7cm, minimum height=0.55cm,
                      font=\scriptsize},
    lbl/.style={font=\scriptsize, inner sep=1.5pt},
    arrow/.style={->, thick, >=Stealth},
]

% === Nodes ===
\node[basenode, minimum width=1.4cm] (csi) at (0,  0)    {CSI Est.};
\node[basenode, minimum width=1.4cm] (det) at (0, -1.5) {Soft MIMO\\ Detector};
\node[smallnode]                     (deint) at (2.5, -1.5) {$\Pi^{-1}$};
\node[basenode, minimum width=1.1cm] (dec) at (4.2, -1.5) {Decoder};
\node[smallnode]                     (int) at (2.5, -2.7) {$\Pi$};

% === Inputs ===
\draw[->, thick] (-1.6,  0)    -- node[above, lbl] {$\bm{Z}, \bm{\Phi}$} (csi.west);
\draw[->, thick] (-1.6, -1.5) -- node[above, lbl] {$\bm{Y}$} (det.west);

% === CSI to Detector ===
\draw[->, thick] (csi.south) -- node[right, lbl, pos=0.4] {$\hat{\bm{H}}, \hat{\sigma}^2$} (det.north);

% === Forward path ===
\draw[->, thick] (det.east)   -- node[above, lbl] {$\bm{L}_{\mathrm{e}}$} (deint.west);
\draw[->, thick] (deint.east) -- (dec.west);
\draw[->, thick] (dec.east)   -- node[above, lbl] {$\hat{\bm{a}}$} ++(0.8, 0);

% === Feedback: Decoder -> Interleaver -> Detector ===
\draw[arrow] (dec.south) |- (int.east);
\draw[arrow] (int.west) -| node[above,lbl, pos=0.3] {$\bm{L}_{\mathrm{e}}^{\mathrm{dec}}$} (det.south);

% === Dashed feedback: Decoder -> CSI Est. (data-aided update) ===
\draw[->, thick, dashed, black!55] (dec.north) -- (4.2, 0) -- (csi.east);
\node[lbl, text=black!55] at (2.1, 0.25) {\textit{CSI update}};

\end{tikzpicture}
        \label{fig:turbo_receiver_conv}
    }
    \hfill
    \subfigure[]{
        % turbo_receiver_icl.tex  (restyled)
\begin{tikzpicture}[font=\fontsize{8pt}{9pt}\selectfont,
    basenode/.style={draw, rounded corners=2pt, thick, align=center,
                     fill=white, minimum height=0.8cm},
    smallnode/.style={draw, rounded corners=2pt, thick, align=center,
                      fill=white, minimum width=0.7cm, minimum height=0.55cm,
                      font=\scriptsize},
    lbl/.style={font=\scriptsize, inner sep=1.5pt},
    arrow/.style={->, thick, >=Stealth},
]

% === Nodes ===
\node[basenode, minimum width=1.6cm, minimum height=0.8cm,fill=green!10, fill opacity=0.6,text opacity=1] (det) at (0, -1.5) {ICL Soft\\Detector};
\node[smallnode] (deint) at (2.5, -1.5) {$\Pi^{-1}$};
\node[basenode, minimum width=1.1cm] (dec) at (4.2, -1.5) {Decoder};
\node[smallnode] (int) at (2.5, -2.7) {$\Pi$};

% === Inputs ===
\draw[arrow] (0, -0.5) -- node[right, lbl] {$\bm{Z}, \bm{\Phi}$} (det.north);
\node[lbl] at (0, -0.3) {pilots};
\draw[arrow] (-1.6, -1.5) -- node[above, lbl] {$\bm{Y}$} (det.west);

% === "No CSI" annotation ===
\node[font=\fontsize{6.5pt}{7.5pt}\selectfont, text=black!40] at (-1.3, -0.7) {no $\hat{\bm{H}}$};

% === Forward path ===
\draw[arrow] (det.east) -- node[above, lbl] {$\bm{L}_{\mathrm{e}}$} (deint.west);
\draw[arrow] (deint.east) -- (dec.west);
\draw[arrow] (dec.east) -- node[above, lbl] {$\hat{\bm{a}}$} ++(0.8, 0);

% === Feedback: Decoder -> Interleaver -> Detector ===
\draw[arrow] (dec.south) |-  (int.east);
\draw[arrow] (int.west) -| node[above,lbl, pos=0.3] {$\bm{L}_{\mathrm{e}}^{\mathrm{dec}}$} (det.south);

\end{tikzpicture}
        \label{fig:turbo_receiver_icl}
    }
    \caption{Block diagrams of a turbo receiver: (a) with a conventional soft detector assisted by estimated CSI, and (b) with an ICL-based soft detector that takes the pilots directly as input.}
    \label{fig:turbo_receivers}
    \vspace{-5mm}
\end{figure}

This paper makes the following contributions:

\begin{itemize}[leftmargin=*,itemsep=2pt,topsep=2pt]
\item \emph{ICL-based soft-input soft-output detection.} We develop EXT-ICLNet, the first ICL-based soft-input soft-output detector for MIMO turbo receivers. In a single forward pass over the pilots and the received data vectors, EXT-ICLNet returns the posterior estimates of all data symbols, from which the bit-level extrinsic LLRs are obtained. EXT-ICLNet accommodates different pilot lengths and quantization resolutions without the need for retraining.

\item \emph{Log-domain prior skip for extrinsic output.} To make EXT-ICLNet compatible with existing turbo receivers, we introduce a log-domain prior skip connection that separates the a priori contribution from the information learned by the network. Specifically, the skip connection adds the a priori log-probabilities of each stream directly to the network output, allowing EXT-ICLNet to learn only the update provided by the received signal and pilots. We show that this architecture enables EXT-ICLNet to approximate extrinsic LLRs using single-stage posterior training, without the specialized training procedures required by existing neural turbo receivers~\cite{zhou2023graph,koike2020neural}.

% \item \emph{Numerical evaluation.} \pp{PP: Numerical evaluation is not a contribution. We can just mention after the contributions how the evaluation has been done.}
% Extensive empirical evaluation shows that EXT-ICLNet achieves a lower bit error rate (BER) than baselines with estimated CSI at low-to-medium signal-to-noise ratios (SNRs), and than EP-based baselines with perfect CSI at high SNRs. The results are further validated by comparing the extrinsic information transfer (EXIT) charts~\cite{ten2001convergence} of EXT-ICLNet, EXT-GEPNet, and neural detector variants without the log-domain prior skip, demonstrating how EXT-ICLNet provides a more favorable transfer of a priori and extrinsic information. 
\end{itemize}

We evaluate EXT-ICLNet in a coded $4\times4$ MIMO system with a quantized front end, and compare it with model-based and neural soft-input soft-output detectors under both perfect and estimated CSI. EXT-ICLNet achieves a lower bit error rate (BER) than the estimated-CSI baselines at low-to-medium signal-to-noise ratios (SNRs) and than the perfect-CSI EP-based baselines at high SNRs. The extrinsic information transfer (EXIT) charts~\cite{ten2001convergence} further show that the EXIT curve of EXT-ICLNet increases monotonically over the evaluated range, whereas those of EXT-GEPNet and of the posterior-output variant without the prior skip decrease at high a priori mutual information.

The rest of the paper is organized as follows. Sec.~\ref{se_model} presents the system model and turbo receiver framework. Sec.~\ref{se:implementation} develops EXT-ICLNet and its theoretical properties. Sec.~\ref{se:results} presents numerical results, and Sec.~\ref{se:conclusion} concludes the paper.

\section{System Model and Turbo Equalization}\label{se_model}

\subsection{Coded MIMO Transmission with Quantized Receiver}\label{subse:transmission}

As shown in Fig.~\ref{fig:system_model_1}, we consider an $N_t \times N_r$ MIMO system with spatial multiplexing and a quantized front end. This setting applies to receivers with low-resolution ADCs in massive MIMO and millimeter-wave systems~\cite{liu2019low,choi2016near,mo2015capacity,horvath2025iterative}, and to fronthaul interfaces that carry quantized baseband samples, as in O-RAN~\cite{oran_cus}. The source generates an information bit sequence $\bm{a} \in \{0,1\}^K$, which is encoded by a forward error correction (FEC) code and subsequently interleaved to obtain the coded bit sequence $\bm{c} \in \{0,1\}^{N_c}$. The interleaved bits are partitioned into groups of $\log_2 M$ bits and mapped to an $M$-ary quadrature amplitude modulation (QAM) constellation $\mathcal{S}$.

The resulting symbols are spatially multiplexed into $T = N_c/(N_t\log_2 M)$ transmit vectors $\bm{x}_i = [x^{(1)}_i, \dots, x^{(N_t)}_i]^\top \in \mathcal{S}^{N_t}$ for $i = 1, \dots, T$, which form the transmitted symbol matrix $\bm{X} = [\bm{x}_1, \dots, \bm{x}_T]$. Before data transmission, the transmitter sends $T_P$ pilot vectors $\{\bm{\phi}_j\}_{j=1}^{T_P}$ with $\bm{\phi}_j \in \mathcal{S}^{N_t}$. The average transmit power is normalized so that $\mathbb{E}[\|\bm{x}_i\|^2] = \mathbb{E}[\|\bm{\phi}_j\|^2] = N_t$.

Under a block-fading model in which the channel is constant over each transmission block~\cite{sellathurai2002turbo}, the received signals for the pilot and data phases are
\begin{subequations}\label{eq:received_signals}
\begin{align}
    \tilde{\bm{z}}_j &= \bm{H} \bm{\phi}_j + \bm{n}'_j, \quad j = 1, \dots, T_P, \\
    \tilde{\bm{y}}_i &= \bm{H} \bm{x}_i + \bm{n}_i, \quad i = 1, \dots, T,
\end{align}
\end{subequations}
where $\bm{H} \in \mathbb{C}^{N_r \times N_t}$ is the channel matrix and $\bm{n}'_j, \bm{n}_i \sim \mathcal{CN}(\bm{0}, \sigma^2 \bm{I})$ are i.i.d.\ additive white Gaussian noise (AWGN) vectors.

The received signals are then quantized by a $B$-bit uniform mid-rise quantizer $Q_B(\cdot)$ with fixed clipping range $[l_{\min}, l_{\max}]$, applied independently to the in-phase and quadrature components, as
\begin{subequations}\label{eq:quantized_signals}
\begin{align}
    \bm{z}_j &= Q_B(\tilde{\bm{z}}_j), \quad j = 1, \dots, T_P, \label{eq_link_pilots}\\
    \bm{y}_i &= Q_B(\tilde{\bm{y}}_i), \quad i = 1, \dots, T. \label{eq_link_symbols}
\end{align}
\end{subequations}
We denote the transmitted and received pilot matrices by $\bm{\Phi} = [\bm{\phi}_1, \dots, \bm{\phi}_{T_P}]$ and $\bm{Z} = [\bm{z}_1, \dots, \bm{z}_{T_P}]$, respectively, and the received data matrix corresponding to $\bm{X}$ by $\bm{Y} = [\bm{y}_1, \dots, \bm{y}_T]$. Under the assumed model, the relation between these quantities is parameterized by the parameters $\tau = \{\bm{H}, \sigma^2, B\}$, comprising the channel matrix, noise variance, and quantization resolution, where $B$ is assumed to be known at the receiver.

\subsection{Soft Detection with Estimated CSI}\label{subse:csi_and_det}

To recover the transmitted symbols $\bm{X}$ from the received signal $\bm{Y}$, conventional receivers first estimate the unknown channel parameters $\{\bm{H}, \sigma^2\}$ from the received pilots $\bm{Z}$ and then perform soft detection with the estimates~\cite{tuchler2002minimum,studer2011asic}.

\subsubsection{Channel Estimation}\label{subsubse:csi}
Given received and transmitted pilots $\bm{Z}$ and $\bm{\Phi}$, the channel matrix $\bm{H}$ can be estimated using least squares (LS) or minimum mean square error (MMSE) estimation, and the noise-variance estimate $\hat{\sigma}^2$ is obtained from the corresponding pilot residuals. In iterative receivers, the channel estimate $\hat{\bm{H}}$ can be refined across turbo iterations by decision-directed RLS updates that exploit the decoded soft symbols~\cite{shiao2008combined,zhang2018soft,da2021adaptive}. After each update, the noise-variance estimate $\hat{\sigma}^2$ is recomputed from the pilot residuals using the current channel estimate $\hat{\bm{H}}$. At each iteration, the refined estimates $(\hat{\bm{H}}, \hat{\sigma}^2)$ are then passed to the soft detector.

\subsubsection{Soft Detection}\label{subsubse:soft_det}
Given the quantized data $\bm{y}_i$, the CSI estimates $(\hat{\bm{H}}, \hat{\sigma}^2)$, and the per-stream symbol priors provided by the decoder (Sec.~\ref{subse:turbo_framework}), $\bm{p}_i \coloneqq (\bm{p}_i^{(1)}, \ldots, \bm{p}_i^{(N_t)})$, where each $\bm{p}_i^{(n)} \coloneqq [p_{i,m}^{(n)}]_{m=1}^{M}$ is the a priori probability mass function (PMF) of stream $n$ and $p_{i,m}^{(n)}$ is the a priori probability that the decoder feedback assigns to $x_i^{(n)} = s_m$ through~\eqref{eq_llr_2_probs}, a conventional soft detector
% is the prior probability mass function (PMF) of stream $n$ with $p_{i,m}^{(n)} \coloneqq P(x_i^{(n)} = s_m)$ {\color{blue}[Written like this it seems like this is the ground-truth symbol probability, while $\bm{p}_i^{(n)}$ is the decoder estimated one no?]}, a conventional soft detector
returns the per-stream posterior estimate $\hat{\bm{q}}_i^{(n)} \coloneqq [\hat{q}_{i,m}^{(n)}]_{m=1}^{M}$, with the $m$-th entry approximating the posterior $P(x_i^{(n)} = s_m \mid \bm{y}_i, \hat{\bm{H}}, \hat{\sigma}^2, \bm{p}_i)$. We refer to this posterior as the \emph{plug-in posterior}, because the estimates $(\hat{\bm{H}}, \hat{\sigma}^2)$ replace the true parameters $(\bm{H}, \sigma^2)$ in the observation model.

Exact maximum a posteriori (MAP) detection requires marginalization over the joint discrete constellation under the nonlinear quantization mapping $Q_B(\cdot)$, with complexity that grows exponentially in the number of transmit streams. Practical detectors such as LMMSE-PIC~\cite{tuchler2002minimum,studer2011asic} and EP-based receivers~\cite{santos2018turbo} use a linear-Gaussian residual model, while neural detectors~\cite{he2020model,zhou2023graph} use a categorical output head to approximate this solution. In all cases, the detection performance is largely influenced by the quality of channel estimates.

\begin{figure}[t]
    \centering
    \subfigure[]{
        % External view of the ICL-based soft detector.
% Requires tikz libraries: arrows.meta, decorations.pathreplacing, calc
\begin{tikzpicture}[
    >={Latex[length=1.8mm]},
    font=\footnotesize,
    detector/.style={draw, rounded corners=3pt, fill=green!10, fill opacity=0.6,
                     text opacity=1, align=center, thick,
                     minimum width=2cm, minimum height=1.8cm, font=\small},
    tok/.style={anchor=west, inner sep=1.5pt}
]

% Prompt tokens (pilot pairs followed by data query)
% \node[tok] (z1) at (0,  1.0) {$(\bm{z}_1, \bm{\phi}_1)$};
% \node[tok] (z2) at (0,  0.5) {$(\bm{z}_2, \bm{\phi}_2)$};
% \node[tok] (zd) at (0,  0.0) {$\vdots$};
% \node[tok] (zT) at (0, -0.5) {$(\bm{z}_{T_P}, \bm{\phi}_{T_P})$};
% \node[tok] (yi) at (0, -1.1) {$[\bm{y}_i, \bm{p}_i]$};

\node[tok] (p) at  (0,  0.0) {$\mathcal{P}_i=\Big(\bm{z}_1, \bm{\phi}_1,\cdots,\bm{z}_{T_P}, \bm{\phi}_{T_P},[\bm{y}_i, \bm{p}_i]\Big)$};

% Right-side brace grouping the prompt as P_i
% 'mirror' flips the curve direction so the brace opens toward the tokens on the left
\draw[decorate, decoration={brace, amplitude=4pt, mirror}] (0.9, -0.3) -- (3.7, -0.3);
\node (Pilab) at (2.3, -0.7) {Context $\mathcal{C}$};

\draw[decorate, decoration={brace, amplitude=4pt, mirror}] (3.8, -0.3) -- (4.8, -0.3);
\node (Pilab) at (4.3, -0.7) {Query};

% Detector box
\node[detector] (det) at (6.5, 0) {ICL Soft\\[1pt] Detector\\[1pt] $\hat{q}_\theta$};

% Input arrow
\draw[->, thick] (p.east) -- (det.west);

% Per-stream output labels
\node[tok] (q1)  at (7.8,  0.70) {$\hat{\bm{q}}_i^{(1)}$};
\node[tok] (q2)  at (7.8,  0.25) {$\hat{\bm{q}}_i^{(2)}$};
\node[tok] (qd)  at (7.8, -0.20) {$\vdots$};
\node[tok] (qNt) at (7.8, -0.70) {$\hat{\bm{q}}_i^{(N_t)}$};

% Output arrows
\draw[->, thick] (det.east |- q1)  -- (q1.west);
\draw[->, thick] (det.east |- q2)  -- (q2.west);
\draw[->, thick] (det.east |- qNt) -- (qNt.west);

\end{tikzpicture}
        \label{fig:arch_external}
    }
    \hfill
    \subfigure[]{
        % Internal view: the Transformer backbone and the log(.) box share the same x
% position on the main vertical axis. Wire-hop convention: vertical residual
% paths pass UNDER horizontal backbone arrows at crossings, indicated by small
% white gaps on the residuals.
% Requires tikz libraries: arrows.meta, positioning, fit, backgrounds, calc
\begin{tikzpicture}[
    >={Latex[length=1.8mm]},
    font=\footnotesize,
    backbone/.style={draw, rounded corners=2pt, thick, align=center,
                     fill=white, minimum width=1.8cm, minimum height=1.5cm,fill=myred!10},
    softmax/.style={draw, rounded corners=2pt, thick, fill=white,
                    minimum width=1cm, minimum height=1.5cm},
    logf/.style={draw, rounded corners=2pt, thick, align=center, fill=white,
                 minimum width=1cm, minimum height=1.2cm},
    adder/.style={draw, circle, thick, fill=white, inner sep=0pt, minimum size=0.48cm},
    branch/.style={circle, fill=black, inner sep=0pt, minimum size=3.5pt},
    wrapper/.style={draw, rounded corners=5pt, fill=green!10, fill opacity=0.6,
                    text opacity=1, thick},
    lbl/.style={inner sep=1.5pt},
    slbl/.style={font=\scriptsize, inner sep=1pt},
    dots/.style={font=\footnotesize, inner sep=0pt}
]

% === Nodes ===
\node[lbl]      (Pin) at (0.3,    0)    {$\mathcal{P}_i$};
\node[branch]   (br)  at (1.0,  0)    {};

\node[backbone] (bb)  at (2.3,  0)    {Transformer\\ Backbone};
\node[logf]     (logn) at (2.3, -1.6) {$\ln(\cdot)$};

\node[adder] (add1) at (4.2,  0.6) {\small $+$};
\node[adder] (add2) at (4.65,  0.1) {\small $+$};
\node[adder] (add3) at (5.3, -0.6) {\small $+$};

\node[softmax] (sm) at (7.0, 0) {softmax};

\node[lbl,font=\scriptsize] (q1) at (8.5,  0.6) {$\hat{\bm{q}}_{i}^{(1)}$};
\node[lbl,font=\scriptsize] (q2) at (8.5,  0.2) {$\hat{\bm{q}}_{i}^{(2)}$};
\node[dots] at (7.7, -0.1) {$\vdots$};
\node[lbl,font=\scriptsize] (q3) at (8.6, -0.6) {$\hat{\bm{q}}_{i}^{(N_t)}$};

\node[dots] at (5.0, -0.1) {$\vdots$};

% === Wrapper ===
\begin{scope}[on background layer]
    \node[wrapper, fit=(br) (bb) (logn) (add1) (add3) (sm), inner sep=8pt] (wrap) {};
\end{scope}

% === East-side log box ports ===
\coordinate (lne1) at (2.8, -1.05);
\coordinate (lne2) at (2.8, -1.45);
\node[dots] at (2.9, -1.65) {$\vdots$};
\coordinate (lne3) at (2.8, -2.05);

\node[font=\scriptsize] at (3.9, -1.25) {$\ln \bm{p}_i^{(1)}$};
\node[font=\scriptsize] at (3.9, -1.65) {$\ln \bm{p}_i^{(2)}$};
\node[font=\scriptsize] at (4.0, -2.25) {$\ln \bm{p}_i^{(N_t)}$};

% === Main path: input -> branch -> backbone ===
\draw[->, thick] (Pin.east) -- (br);
\draw[->, thick] (br) -- node[pos=0.5, above] {$\mathcal{P}_i$} (bb.west);

% === Residual branch: br -> log west, labeled with the full prior p_i ===
\draw[->, thick] (br) |- node[pos=0.8, above] {$\bm{p}_i$} (logn.west);

% === Residuals from log box up to adders (drawn FIRST, will be hopped over) ===
\draw[->, thick] (lne1) -| (add1.south);
\draw[->, thick] (lne2) -| (add2.south);
\draw[->, thick] (lne3) -| (add3.south);

% === White hops at crossings ===
% Convention: vertical residuals pass UNDER horizontal backbone arrows.
% Mask color matches the wrapper background (green!10 with 0.6 opacity over white,
% which renders as a pale green). We use green!6 as a close visual match.
\definecolor{wrapperbg}{rgb}{0.93, 0.97, 0.93}  % approx of green!10 @ 60% opacity on white
\draw[wrapperbg, line width=2.6pt] (4.2, 0.15)  -- (4.2, -0.02);
\draw[wrapperbg, line width=2.6pt] (4.2, -0.48) -- (4.2, -0.72);
\draw[wrapperbg, line width=2.6pt] (4.65, -0.48) -- (4.65, -0.72);

% === Backbone -> three adders (drawn AFTER masks, appear on top) ===
\draw[->, thick] (bb.east |- add1) -- (add1.west);
\draw[->, thick] (bb.east |- add2) -- (add2.west);
\draw[->, thick] (bb.east |- add3) -- (add3.west);

\node[font=\scriptsize] at (3.5, 0.9) {$\bm{l}_{\mathrm{e},i}^{(1)}$};
\node[font=\scriptsize] at (3.5, 0.3) {$\bm{l}_{\mathrm{e},i}^{(2)}$};
\node[font=\scriptsize] at (3.6, -0.4) {$\bm{l}_{\mathrm{e},i}^{(N_t)}$};

% === Adders -> shared softmax ===
\draw[->, thick] (add1.east) -- (sm.west |- add1);
\draw[->, thick] (add2.east) -- (sm.west |- add2);
\draw[->, thick] (add3.east) -- (sm.west |- add3);

\node[font=\scriptsize] at (6.1, 0.9) {$\bm{l}_{i}^{(1)}$};
\node[font=\scriptsize] at (6.1, 0.3) {$\bm{l}_{i}^{(2)}$};
\node[font=\scriptsize] at (6.2, -0.4) {$\bm{l}_{i}^{(N_t)}$};

% === Softmax -> per-stream posterior outputs ===
\draw[->, thick] (sm.east |- q1) -- (q1.west);
\draw[->, thick] (sm.east |- q2) -- (q2.west);
\draw[->, thick] (sm.east |- q3) -- (q3.west);

\end{tikzpicture}
        \label{fig:arch_internal}
    }
    \caption{EXT-ICLNet architecture: (a)~the prompt $\mathcal{P}_i$ and the per-stream posterior estimates $\hat{\bm{q}}_i^{(n)}$; (b)~the shared Transformer backbone, stream-specific output heads, and log-domain prior skip.}
    \label{fig:architecture}
\end{figure}

\subsection{Iterative Detection and Decoding in Turbo Receivers}\label{subse:turbo_framework}

As shown in Fig.~\ref{fig:turbo_receiver_conv}, the turbo receiver alternates between soft detection, as described in Sec.~\ref{subse:csi_and_det}, and soft decoding, which exploits the code structure to refine the bit-level likelihood ratios. Information is exchanged between these two components through extrinsic LLRs, which capture the information contributed by each component beyond the a priori information received from the other component~\cite{tuchler2011turbo}. More specifically, at each iteration, the decoder-extrinsic LLR vector $\bm{L}_{\mathrm{e}}^{\mathrm{dec}}\coloneqq[L_{\mathrm{e}}^{\mathrm{dec}}(c_1),\ldots,L_{\mathrm{e}}^{\mathrm{dec}}(c_{N_c})]^\top$ from the previous iteration serves as the a priori information of the detector and is initialized to zero at the first iteration. The detector converts these LLRs to per-stream symbol priors under the bit-independence assumption of bit-interleaved coded modulation (BICM)~\cite{caire1998bit}. Specifically, the symbol prior PMF for stream $n$ factorizes as
\begin{equation}\label{eq_llr_2_probs}
    p_{i,m}^{(n)} = \prod_{\kappa=1}^{\log_2 M} P\big(c_{\pi(i,n,\kappa)}=b_{\kappa,m}\big),
\end{equation}
where $b_{\kappa,m}\in\{0,1\}$ is the $\kappa$-th bit of the label of $s_m$, $\pi(i,n,\kappa)=((i-1)N_t+n-1)\log_2 M+\kappa$ is the corresponding index in the interleaved codeword, and $P(c_k=b)=\frac{1}{2}[1+(2b-1)\tanh(L_{\mathrm{e}}^{\mathrm{dec}}(c_k)/2)]$. After detection, the posterior estimate $\hat{\bm{q}}_i^{(n)}$ for stream $n$ is marginalized to the bit-level posterior LLR
\begin{equation}\label{eq_probs_to_llr}
    L\big(c_{\pi(i,n,\kappa)}\mid\hat{\bm{q}}_i^{(n)}\big) = \ln\frac{\sum_{s_m\in\mathcal{S}_{\kappa,1}}\hat{q}_{i,m}^{(n)}}{\sum_{s_m\in\mathcal{S}_{\kappa,0}}\hat{q}_{i,m}^{(n)}},
\end{equation}
where $\mathcal{S}_{\kappa,b}=\{s_m\in\mathcal{S}:b_{\kappa,m}=b\}$. The corresponding detector-extrinsic LLR is obtained by subtracting the prior information from the posterior as
\begin{equation}\label{eq_llr_chn_ext}
    L_{\mathrm{e}}\big(c_{\pi(i,n,\kappa)}\big) \coloneqq L\big(c_{\pi(i,n,\kappa)}\mid\hat{\bm{q}}_i^{(n)}\big) - L_{\mathrm{e}}^{\mathrm{dec}}\big(c_{\pi(i,n,\kappa)}\big),
\end{equation}
and it is clipped to prevent numerical instability~\cite{de2005iterative}. The detector-extrinsic LLRs are collected in the vector $\bm{L}_{\mathrm{e}}\coloneqq[L_{\mathrm{e}}(c_1),\ldots,L_{\mathrm{e}}(c_{N_c})]^\top$ and passed to the decoder. The decoder computes the posterior LLRs of the coded bits from $\bm{L}_{\mathrm{e}}$ leveraging the code structure, and returns its extrinsic LLR vector $\bm{L}_{\mathrm{e}}^{\mathrm{dec}}$, subtracting $\bm{L}_{\mathrm{e}}$ from the posterior LLRs. The receiver iterates until all code parity checks are satisfied or the maximum number of iterations is reached.

\section{Soft Detection via In-Context Learning}\label{se:implementation}
 
The turbo receiver of Sec.~\ref{se_model} has two limitations. The extrinsic LLRs in~\eqref{eq_llr_chn_ext} inherit the errors of the channel estimates $(\hat{\bm{H}}, \hat{\sigma}^2)$, and, for a neural detector, the subtraction in~\eqref{eq_llr_chn_ext} is only effective if the posterior LLRs are computed exactly~\cite{yang2013approximate,yang2013nominal,zhou2023graph,clausius2025joint}. This section addresses both limitations with a single-stage neural detector based on ICL~\cite{muller2022transformers}. As shown in Fig.~\ref{fig:architecture}, the detector maps the pilot sequence, the a priori information, and the received data signal directly to the symbol posterior through a log-domain prior skip that ensures the direct learning of extrinsic LLRs.

\subsection{Formulation of ICL-Based Soft Detection}\label{subse:icl_eq_general}

We collect the $T_P$ pairs of received and transmitted pilots to form a context $\mathcal{C} \coloneqq (\bm{z}_1, \bm{\phi}_1, \ldots, \bm{z}_{T_P}, \bm{\phi}_{T_P})$.
As shown in Fig.~\ref{fig:architecture}(a), for each received data vector $\bm{y}_i$ with prior $\bm{p}_i$, the detector combines the context and the query $[\bm{y}_i, \bm{p}_i]$ to form the input prompt
\begin{equation}\label{eq_prompt_inference}
    \mathcal{P}_i = \big(\, \mathcal{C},\, \underbrace{[\bm{y}_i, \bm{p}_i]}_{\text{query}} \,\big),
\end{equation}
which is mapped to the per-stream posterior estimates $(\hat{\bm{q}}_i^{(n)})_{n=1}^{N_t}$. The $m$-th entry of $\hat{\bm{q}}_i^{(n)}$ can be written as
\begin{equation}\label{eq_factorized_posterior}
    \hat{q}_{i,m}^{(n)} = \hat{q}_\theta^{(n)}(s_m \mid \mathcal{C}, \bm{y}_i, \bm{p}_i),
\end{equation}
where $\hat{q}_\theta^{(n)}(\cdot \mid \mathcal{C}, \bm{y}_i, \bm{p}_i)$ is a parameterized function with $\theta$ collecting the detector's parameters. We write $\hat{q}_\theta=(\hat{q}_\theta^{(1)},\dots,\hat{q}_\theta^{(N_t)})$ for the full per-stream collection. 
 
The detector parameters $\theta$ are optimized to minimize the population per-stream cross-entropy
\begin{equation}\label{eq_expected_loss}
    \mathcal{L}_\theta \coloneqq \mathbb{E}_{\tau, \bm{p}_i, \bm{x}_i, \bm{y}_i, \mathcal{C}}\!\left[-\sum_{n=1}^{N_t} \ln \hat{q}_\theta^{(n)}(x_i^{(n)} \mid \mathcal{C}, \bm{y}_i, \bm{p}_i)\right],
\end{equation}
with minimizer $\theta^* = \arg\min_\theta \mathcal{L}_\theta$. The expectation in~\eqref{eq_expected_loss} is taken over the following generative model. The system parameters follow $\tau \sim \mathcal{D}_\tau$, where $\mathcal{D}_\tau$ is the distribution of the channel matrix, noise variance, and quantization resolution, and the priors follow $\bm{p}_i \sim \mathcal{D}_p$, where $\mathcal{D}_p$ is the distribution of the a priori PMFs that the decoder provides across turbo iterations. Following the ICL terminology, we refer to $\tau$ as the task and to $\mathcal{D}_\tau$ as the task distribution. Given the prior $\bm{p}_i$, each symbol $x_i^{(n)}$ is drawn from $\bm{p}_i^{(n)}$ independently across streams~\cite{ten2001convergence}. Given the pilots $\{\bm{\phi}_j\}$ and the data symbols $\bm{x}_i$, the observations $\{\bm{z}_j\}$ and $\bm{y}_i$ follow~\eqref{eq:received_signals} and~\eqref{eq:quantized_signals}, respectively.
% with minimizer $\theta^* = \arg\min_\theta \mathcal{L}_\theta$, where the system parameters $\tau$ and the priors follow the distributions $\tau \sim \mathcal{D}_\tau$ and $\bm{p}_i \sim \mathcal{D}_p$, respectively. {\color{blue} [Where are $\mathcal{D}_p$ and $\mathcal{D}_\tau$ defined?]} Following the ICL terminology, we refer to $\tau$ as the task and to $\mathcal{D}_\tau$ as the task distribution. For given pilots $\{\bm{\phi}_j\in\mathcal{S}^{N_t}\}$ and data symbols $\bm{x}_i \sim \bm{p}_i$ {\color{blue} [Why is the data symbol $\bm{x}_i$ distributed as the prior?]}, the observations $\{\bm{z}_j\}$ and $\bm{y}_i$ follow~\eqref{eq:received_signals} and~\eqref{eq:quantized_signals}, respectively.

\begin{figure}[t]
    \centering
    % training_prompt_gen.tex
% Requires tikz libraries: arrows.meta, positioning, calc
\begin{tikzpicture}[
    >={Latex[length=1.6mm, width=1.4mm]},
    font=\footnotesize,
    line width=0.5pt,
    block/.style={draw, align=center, fill=white, inner sep=4pt,
                  minimum height=1cm},
    data/.style={draw, thick, align=center, fill=white, inner sep=4pt,
                 rounded corners=2pt},
    query/.style={data, fill=mygreen!15},
    label/.style={data, fill=myyellow!20},
    qtok/.style={data, fill=mygreen!15, inner sep=1.5pt, font=\scriptsize,
                 minimum height=0.45cm},
    ltok/.style={data, fill=myyellow!20, inner sep=1.5pt, font=\scriptsize,
                 minimum height=0.45cm},
    arrow/.style={->, line width=0.5pt},
    edgelbl/.style={inner sep=1.5pt, font=\scriptsize},
    indep/.style={text=black!55, inner sep=1.5pt, font=\scriptsize}
]

% === Column anchors ===
\def\colL{0}
\def\colM{2.4}
\def\colR{4.8}

% === Top row: task and prior sampling ===
\node[block] (tau) at (\colL, 0)
    {Sample task\\ $\tau \sim \mathcal{D}_{\tau}^{\text{train}}$};

\node[block] (dir) at (\colR, 0)
    {Sample $\bm{p}_t^{(n)} \sim \mathrm{Dir}(\alpha\mathbf{1}_M)$\\[1pt]
     $\alpha \sim \mathcal{D}_{\alpha}$\\[1pt]
     $n{=}1,\ldots,N_t$;\; $t{=}1,\ldots,T_{\max}$};

% === Middle row: observation and symbol generation ===
\node[block] (y) at (\colL, -2)
    {Generate\\ $\bm{y}_t = Q_B(\bm{H}\bm{x}_t + \bm{n}_t)$};

\node[block] (x) at (\colR, -2)
    {Draw $x_t^{(n)} \sim \bm{p}_t^{(n)}$\\ for each stream and position};

% === Query/label tokens ===
\node[query, minimum width=1.7cm] (qbox) at (\colL, -3.5)
    {Query:\; $[\bm{y}_t,\,\bm{p}_t]$};

\node[label, minimum width=1.7cm] (lbox) at (\colR, -3.5)
    {Label:\; $\bm{x}_t$};

% === Interleave block ===
\node[block, minimum width=3.5cm,minimum height=0.5cm] (prompt) at (\colM, -4.7)
    {Form $\mathcal{P}_{\text{train}}$~\eqref{eq_prompt_train}};

% === Independence marker ===
\draw[black!30] ($(tau.east)+(0.12,0)$) -- ($(dir.west)+(-0.12,0)$)
    node[midway, fill=white, indep] {independent};

% === Vertical edges ===
\draw[arrow] (tau) -- node[left, edgelbl] {$\bm{H},\sigma^2,B$} (y);
\draw[arrow] (dir) -- node[right, edgelbl] {$\bm{p}_t$} (x);

% === p_t residual: drawn first so x_t can hop over it ===
\def\pjx{2.5}
\draw[arrow, rounded corners=3pt]
    ($(dir.south)+(-0.3,0)$) -- ++(0,-0.2)
    -| (\pjx, -2)
    node[edgelbl, anchor=west, pos=0.75]
        {$\bm{p}_t$}
    -- (\pjx, -3.5) -- (qbox.east);

% === Cross edge x_t -> y with a wire-hop over p_t ===
\draw[white, line width=2.6pt] (\pjx, -1.90) -- (\pjx, -2.10);
\draw[arrow] (x.west) -- node[above, edgelbl, pos=0.7] {$\bm{x}_t$} (y.east);

% === Edges into query/label ===
\draw[arrow] (y) -- (qbox);
\draw[arrow] (x) -- (lbox);

% === Edges into the interleave block ===
\coordinate (pL) at ($(prompt.north)+(-0.8,0)$);
\coordinate (pR) at ($(prompt.north)+(0.8,0)$);
\draw[arrow] (qbox.south) |- ($(pL)+(0,0.30)$) -- (pL);
\draw[arrow] (lbox.south) |- ($(pR)+(0,0.30)$) -- (pR);
\draw[arrow] (prompt.south) -- ++(0,-0.4);

% === Prompt visualization ===
\def\py{-5.9}
\node (ptrain) at (-0.7, \py) {$\mathcal{P}_{\text{train}}=\Big($};
\node[qtok, right=0cm of ptrain]    (q1)   {$[\bm{y}_1,\,\bm{p}_1]$};
\node[ltok, right=0.08cm of q1]     (a1)   {$\bm{x}_1$};
\node[qtok, right=0.08cm of a1]     (q2)   {$[\bm{y}_2,\,\bm{p}_2]$};
\node[ltok, right=0.08cm of q2]     (a2)   {$\bm{x}_2$};
\node[right=0cm of a2]              (dots) {$\cdots$};
\node[qtok, right=0.08cm of dots]   (qT)   {$[\bm{y}_{T_{\max}},\,\bm{p}_{T_{\max}}]$};
\node[ltok, right=0.08cm of qT]     (aT)   {$\bm{x}_{T_{\max}}$};
\node[right=0cm of aT]              (close){$\Big)$};

\end{tikzpicture}
    \caption{Training data generation. For each training prompt, the task $\tau$ and the prior PMFs $\bm{p}_t$ are sampled independently. The symbols $\bm{x}_t$ are drawn from the priors, the received signals $\bm{y}_t$ are generated under the sampled task, and the resulting query-label pairs form the training sequence $\mathcal{P}_{\text{train}}$.}
    \label{fig:data_gen}
\end{figure}

Unlike the conventional detector of Sec.~\ref{subsubse:soft_det}, which conditions on the point estimates $(\hat{\bm{H}}, \hat{\sigma}^2)$ and targets the plug-in posterior, an ICL-based soft detector conditions directly on the pilot context $\mathcal{C}$, without explicit channel estimation, and targets the posterior $\bm{q}_i^{(n)} \coloneqq [q_{i,m}^{(n)}]_{m=1}^{M}$, with entries
\begin{equation}\label{eq_true_posterior}
    q_{i,m}^{(n)} \coloneqq P(x_i^{(n)} = s_m \mid \mathcal{C}, \bm{y}_i, \bm{p}_i).
\end{equation}
Because~\eqref{eq_true_posterior} conditions on the pilots instead of a channel estimate, it accounts for the channel uncertainty that remains after the pilots are observed~\cite{muller2022transformers,xie2022explanation,zecchin2023context,rajagopalan2023transformers,zecchin2026context}.

\subsection{Training Distribution and Posterior Factorization}\label{subse:extrinsic_sep}

% {\color{blue}[I had problems following this section. Wouldn't make sense to explain first that the population loss is unknown since we do not know 1) the channel parameters distribution 2) the distribution of the prior LLRs given by the decoder across iterations. For this reason, we approximate the population objective by positing a training task distribution and sampling $N_{train}$ channel parameters, while for the prior distribution, we model the successive refinement of turbo iteration by creating progressively more concentrated priors using a Dirichlet model? I think providing more intuition behind these choices would be helpful]}

% Each training sequence is composed of $T_{\max}$ query-label pairs, written as
The population loss~\eqref{eq_expected_loss} cannot be evaluated because neither the task distribution $\mathcal{D}_\tau$ nor the prior distribution $\mathcal{D}_p$ are known. The distribution of the decoder feedback, in particular, depends on the code, the SNR, and the turbo iteration. We therefore train on synthetic data drawn from a training task distribution $\mathcal{D}_\tau^{\text{train}}$ and from a family of prior distributions whose concentration ranges from the nearly uniform priors of the first turbo iteration to the nearly deterministic priors of later iterations. Fig.~\ref{fig:data_gen} summarizes the procedure.

Each training sequence is generated under a single task $\tau$ and is composed of $T_{\max}$ query-label pairs, written as
\begin{equation}\label{eq_prompt_train}
    \mathcal{P}_{\text{train}} = \big(
        [\bm{y}_1, \bm{p}_1], \bm{x}_1, \ldots, [\bm{y}_{T_{\max}}, \bm{p}_{T_{\max}}], \bm{x}_{T_{\max}}
    \big).
\end{equation}
% generated following the process in Fig.~\ref{fig:data_gen}. {\color{blue} [This is for a single pilot sequence?]}
The training sequence contains no separate pilot phase. At position $t$, the first $t-1$ query-label pairs act as the context, in the role that the pilot pairs play at inference, so that one sequence provides training examples for every context length up to $T_{\max}-1$.

For each training prompt, we first sample a task $\tau$ from the training task distribution $\mathcal{D}_\tau^{\text{train}}$.
% {\color{blue}[Why not $\mathcal{P}$ for these distributions?]} Because P has been used to denote prompt 
We adopt the product distribution
\begin{equation}
    \mathcal{D}_\tau^{\text{train}} = \mathcal{D}_{\bm{H}}^{\text{train}} \times \mathcal{D}_{\sigma^2}^{\text{train}} \times \mathcal{D}_{B}^{\text{train}},
\end{equation}
where $\mathcal{D}_{\bm{H}}^{\text{train}}$, $\mathcal{D}_{\sigma^2}^{\text{train}}$, and $\mathcal{D}_{B}^{\text{train}}$ are the training distributions for the channel, noise variance, and quantization resolution, respectively. The training channel distribution is $\mathcal{D}_{\bm{H}}^{\text{train}} = \mathcal{U}(\mathcal{H}_{\text{train}})$, drawn uniformly from the fixed channel set
\begin{equation}
    \mathcal{H}_{\text{train}} = \{\bm{H}_k\}_{k=1}^{N_{\text{train}}} \sim \mathcal{D}_{\bm{H}}^{\otimes N_{\text{train}}},
\end{equation}
sampled once prior to training from a channel distribution $\mathcal{D}_{\bm{H}}$, with $N_{\text{train}}$ denoting the number of distinct channel realizations in the fixed training set. The specific distribution $\mathcal{D}_{\bm{H}}$ used to generate this set is given in Sec.~\ref{subse:setup}. The noise variance and quantization resolution, $\sigma^2 \sim \mathcal{D}_{\sigma^2}^{\text{train}}$ and $B \sim \mathcal{D}_{B}^{\text{train}}$, are instead drawn independently for each prompt.
 
The prior PMFs are generated independently of the task $\tau$. To cover the range of decoder feedback encountered across turbo iterations, we draw a concentration parameter $\alpha\sim\mathcal{D}_\alpha$ once per training prompt and use it for all positions and streams in that prompt. Conditioned on $\alpha$, the prior PMF $\bm{p}_t^{(n)}$ for position $t$ and stream $n$ is drawn independently from the symmetric Dirichlet distribution
\begin{equation}
    \bm{p}_t^{(n)} \mid \alpha
    \overset{\mathrm{i.i.d.}}{\sim}
    \mathrm{Dir}(\alpha\bm{1}_M),
\end{equation}
where $\bm{1}_M\in\mathbb{R}^{M}$ denotes the all-ones vector. Smaller values of $\alpha$ produce more concentrated prior PMFs, as in later turbo iterations, whereas larger values produce more diffuse prior PMFs, as in the early iterations. Given the prior PMF $\bm{p}_t^{(n)}$, the transmitted symbol $x_t^{(n)}$ is drawn according to
\begin{equation}
    P\!\left(x_t^{(n)}=s_m\mid\bm{p}_t^{(n)}\right)
    =
    p_{t,m}^{(n)},
    \quad m\in\{1,\ldots,M\}.
\end{equation}
The received signal $\bm{y}_t$ is then generated according to~\eqref{eq_link_symbols} under the sampled task $\tau$. The distribution of the concentration parameter $\alpha$ used in the numerical evaluations is specified in Sec.~\ref{subse:setup}.

For target position $t\in\{2,\ldots,T_{\max}\}$, let
\begin{equation}
    \mathcal{C}_{t}^{\mathrm{train}}
    \coloneqq
    \big(
    [\bm{y}_1,\bm{p}_1],\bm{x}_1,\ldots,
    [\bm{y}_{t-1},\bm{p}_{t-1}],\bm{x}_{t-1}
    \big)
\end{equation}
denote the preceding query-label pairs available as context, and let $q_{t,m}^{(n)} \coloneqq P(x_t^{(n)} = s_m \mid \mathcal{C}_{t}^{\mathrm{train}}, \bm{y}_t, \bm{p}_t)$ denote the corresponding posterior. Following the generative procedure above, Bayes' rule gives
\begin{equation}\label{eq_posterior_factorization}
    q_{t,m}^{(n)} \propto \ell_{t}^{(n)}(s_m) p_{t,m}^{(n)},
\end{equation}
where
\begin{align}\label{eq_marg_lik}
    \ell_{t}^{(n)}(s_m) \coloneqq \sum_{\bm{x}_t^{(\backslash n)}} P\big(\bm{y}_t \mid x_t^{(n)} = s_m, \bm{x}_t^{(\backslash n)}, \mathcal{C}_{t}^{\mathrm{train}}\big)\,\nonumber\\ 
    \cdot P\big(\bm{x}_t^{(\backslash n)} \mid \bm{p}_t^{(\backslash n)}\big)
\end{align}
is the likelihood of the received signal $\bm{y}_t$ under $x_t^{(n)}=s_m$, marginalized over the symbols $\bm{x}_t^{(\backslash n)}$ on the other streams and, through the training context $\mathcal{C}_{t}^{\mathrm{train}}$, over the channel. Here, $(\cdot)^{(\backslash n)}$ denotes the corresponding quantity collected over all streams other than $n$. The same factorization applies at inference with the training context $\mathcal{C}_{t}^{\mathrm{train}}$ replaced by the pilot context $\mathcal{C}$; the corresponding likelihood for data query $i$ and stream $n$ is denoted by $\ell_{i}^{(n)}(s_m)$. 
 
\subsection{Log-Domain Prior Skip}\label{subse:residual_head}

% {\color{blue} [Isn't the role of the skip connection that of adding the prior input to the target output of the model so the model has to learn the difference between the two, which corresponds to the extrinsic information? If so, would it not be clearer explained like this?]}

% Since the likelihood term $\ell_{i}^{(n)}(s_m)$ does not depend on the own-stream prior, we implement~\eqref{eq_posterior_factorization} using the fixed log-domain prior skip shown in Fig.~\ref{fig:architecture}(b). Specifically, let 
By~\eqref{eq_posterior_factorization}, the logarithm of the target posterior is, up to a constant, the sum of the log-prior $\ln p_{i,m}^{(n)}$, which is an input of the detector, and the log-likelihood $\ln \ell_{i}^{(n)}(s_m)$, which does not depend on the own-stream prior and carries the extrinsic information about stream $n$. The log-domain prior skip in Fig.~\ref{fig:architecture}(b) adds the log-prior directly to the output of the network, so that the network is forced to learn only the difference between the target posterior and its prior input, i.e., the log-likelihood term.

Specifically, let $\bm{l}_{\mathrm{e},i}^{(n)} \in \mathbb{R}^{M}$ denote the residual logit vector produced by the Transformer backbone for data query $i$ and stream $n$. The posterior logit vector $\bm{l}_{i}^{(n)} \in \mathbb{R}^{M}$ is obtained as
\begin{equation}\label{eq_logit_decomp}
    \bm{l}_{i}^{(n)}
    =
    \bm{l}_{\mathrm{e},i}^{(n)}
    +
    \ln \bm{p}_{i}^{(n)},
\end{equation}
where the logarithm is applied element-wise, with the prior probabilities clipped away from zero for numerical stability. Applying the softmax function gives the posterior estimate
\begin{equation}\label{eq_residual_posterior}
    \hat{q}_{i,m}^{(n)}
    =
    \frac{
        \exp\!\left(
            l_{\mathrm{e},i,m}^{(n)}
            +
            \ln p_{i,m}^{(n)}
        \right)
    }{
        \displaystyle
        \sum_{m'=1}^{M}
        \exp\!\left(
            l_{\mathrm{e},i,m'}^{(n)}
            +
            \ln p_{i,m'}^{(n)}
        \right)
    },
    \quad m \in \{1,\ldots,M\}.
\end{equation}

Comparing~\eqref{eq_posterior_factorization} and~\eqref{eq_residual_posterior}, if the model can represent the population-optimal posterior, the corresponding residual logits can be written as
\begin{equation}\label{eq_residual_at_optimum}
    l_{\mathrm{e},i,m}^{(n)}\!\left(\theta^{*}\right)
    =
    \ln \ell_{i}^{(n)}(s_m)
    +
    c_{i}^{(n)},
\end{equation}
where $c_{i}^{(n)}$ is a scalar that does not depend on the constellation index $m$ and is therefore removed by the softmax normalization.

For comparison, consider a posterior-output network
$\hat{\bm{q}}_{\theta}^{(n)}
=
\operatorname{softmax}(\bm{g}_{\theta}^{(n)})$,
where $\bm{g}_{\theta}^{(n)} \in \mathbb{R}^{M}$ denotes the logits produced directly by the network. In the posterior-output variant of EXT-ICLNet evaluated in Sec.~\ref{subse:exit_robustness}, the same backbone produces $\bm{g}_{\theta}^{(n)}$ without the prior skip in~\eqref{eq_logit_decomp}, and the extrinsic information is obtained by removing the prior from the posterior estimate as $\hat{\ell}_{i}^{(n)}(s_m)\propto\hat{q}_{i,m}^{(n)}/p_{i,m}^{(n)}$. Similar logit-addition structures have been used with static class priors in long-tailed classification~\cite{menon2021longtail} and in neural turbo equalization~\cite{koike2020neural}; here, the prior $\bm{p}_{i}^{(n)}$ varies across data symbols and turbo iterations.

\subsection{Implementation}\label{subse:architecture}

In this subsection, we detail a specific implementation of the proposed ICL-based detector realized via a Transformer model. We describe how the prompt is represented using the tokens and detail the training and inference procedures.

Each query or label is represented by the token vector
\begin{equation}\label{eq_tokenization}
    \bm{t} = \left[\Re\!\left(\bm{r}\right)^{\top}, \Im\!\left(\bm{r}\right)^{\top}, \bm{v}^{\top}\right]^{\top} \in \mathbb{R}^{2\max(N_r,N_t)+MN_t},
\end{equation}
where the signal field $\bm{r}\in\mathbb{C}^{\max(N_r,N_t)}$ and the prior field $\bm{v}\in\mathbb{R}^{MN_t}$ depend on the token type. The signal field is zero-padded when its original dimension is smaller than $\max(N_r,N_t)$.

For a data-query token, the signal field $\bm{r}$ contains the quantized received signal $\bm{y}_{i}$, while the prior field $\bm{v}$ contains the vector obtained by concatenating the per-stream prior PMFs $\bm{p}_{i}^{(1)},\ldots,\bm{p}_{i}^{(N_t)}$. For a pilot-query token, the signal field contains the quantized pilot observation $\bm{z}_{j}$, while the prior field contains the concatenation of $N_t$ uniform PMFs, each assigning probability $1/M$ to every constellation point. For a data-label token used during training, the signal field contains the transmitted symbol vector $\bm{x}_{t}$ and the prior field is the all-zero vector in $\mathbb{R}^{MN_t}$. Similarly, for a pilot-label token used at inference, the signal field contains the pilot vector $\bm{\phi}_{j}$ and the prior field is the all-zero vector.

\begin{figure}[t]
    \centering
    \subfigure[Training with teacher forcing and causal attention.]{%
        % fig_pipeline_train.tex
% Requires tikz libraries: arrows.meta, calc
\begin{tikzpicture}[
    >={Latex[length=1.6mm, width=1.4mm]},
    font=\footnotesize,
    line width=0.5pt,
    block/.style={draw, align=center, fill=white, inner sep=4pt,
                  rounded corners=2pt},
    data/.style={draw, thick, align=center, fill=white, inner sep=2pt,
                 rounded corners=2pt, minimum height=0.55cm, minimum width=0.5cm},
    query/.style={data, fill=mygreen!15, font=\scriptsize},
    label/.style={data, fill=myyellow!20, font=\scriptsize},
    addnode/.style={draw, thick, circle, minimum size=0.28cm,
                    inner sep=0pt, font=\scriptsize, fill=white},
    arrow/.style={->, line width=0.5pt},
    bus/.style={line width=0.5pt},
    attn/.style={line width=0.3pt, black!48},
    edgelbl/.style={inner sep=1.5pt, font=\scriptsize},
    junction/.style={circle, fill=black, inner sep=0pt, minimum size=2.2pt}
]

% === Layout parameters ===
\def\ax{-3.6}
\def\dx{0.8}
\def\ty{0}
\def\tboxtop{-0.55}
\def\tboxbot{-1.3}
\def\ley{-1.85}
\def\addy{-2.55}
\def\busy{-3.10}
\def\tfpad{0.6}

% Token x-coordinates
\pgfmathsetmacro{\xa}{\ax}
\pgfmathsetmacro{\xb}{\ax+\dx}
\pgfmathsetmacro{\xc}{\ax+2*\dx}
\pgfmathsetmacro{\xd}{\ax+3*\dx}
\pgfmathsetmacro{\xe}{\ax+4*\dx}
\pgfmathsetmacro{\xf}{\ax+5*\dx}
\pgfmathsetmacro{\xdots}{\ax+6*\dx}
\pgfmathsetmacro{\xg}{\ax+7*\dx}
\pgfmathsetmacro{\xh}{\ax+8*\dx}

% === Input tokens ===
\node[query] at (\xa, \ty) {$\bm{y}_1$\\$\bm{p}_1$};
\node[label] at (\xb, \ty) {$\bm{x}_1$};
\node[query] at (\xc, \ty) {$\bm{y}_2$\\$\bm{p}_2$};
\node[label] at (\xd, \ty) {$\bm{x}_2$};
\node[query] at (\xe, \ty) {$\bm{y}_3$\\$\bm{p}_3$};
\node[label] at (\xf, \ty) {$\bm{x}_3$};
\node[edgelbl] at (\xdots, \ty) {$\cdots$};
\node[query] at (\xg, \ty) {$\bm{y}_{T_{\max}}$\\$\bm{p}_{T_{\max}}$};

% === Transformer box ===
\pgfmathsetmacro{\tfl}{\ax-\tfpad}
\pgfmathsetmacro{\tfr}{\xg+\tfpad}
\fill[myred!8, rounded corners=2pt] (\tfl, \tboxtop) rectangle (\tfr, \tboxbot);
\draw[rounded corners=2pt] (\tfl, \tboxtop) rectangle (\tfr, \tboxbot);

% === Input arrows into transformer ===
\foreach \x in {\xa,\xb,\xc,\xd,\xe,\xf,\xg} {
    \draw[arrow] (\x, \ty-0.30) -- (\x, \tboxtop);
}

% === Causal attention at the target positions t = 2, 3, ..., T_max ===
% (the first query-label pair serves as initial context only)
\foreach \src in {\xa,\xb,\xc} {
    \draw[attn] (\src, \tboxtop) -- (\xc, \tboxbot);
}
\foreach \src in {\xa,\xb,\xc,\xd,\xe} {
    \draw[attn] (\src, \tboxtop) -- (\xe, \tboxbot);
}
\foreach \src in {\xa,\xb,\xc,\xd,\xe,\xf,\xg} {
    \draw[attn] (\src, \tboxtop) -- (\xg, \tboxbot);
}

% === Transformer label ===
\node[draw,anchor=west, font=\scriptsize, fill=white, inner sep=1.5pt,align=center]
    at ({\tfr-0.3}, {\tboxtop-0.2}) {Transformer\\(causal mask)};

% === Output residual logits at the target positions ===
\node[edgelbl] at (\xc, \ley) {$\bm{l}_{\mathrm{e},2}^{(n)}$};
\node[edgelbl] at (\xe, \ley) {$\bm{l}_{\mathrm{e},3}^{(n)}$};
\node[edgelbl] at (\xdots, \ley) {$\cdots$};
\node[edgelbl] at (\xg, \ley) {$\bm{l}_{\mathrm{e},T_{\max}}^{(n)}$};

\foreach \x in {\xc,\xe,\xg} {
    \draw[arrow] (\x, \tboxbot) -- (\x, \ley+0.18);
}

% === Adders ===
\node[addnode] at (\xc, \addy) (add2) {$+$};
\node[addnode] at (\xe, \addy) (add3) {$+$};
\node[addnode] at (\xg, \addy) (addL) {$+$};

\foreach \x/\add in {\xc/add2, \xe/add3, \xg/addL} {
    \draw[arrow] (\x, \ley-0.18) -- (\add);
}

% === Prior labels ===
\node[edgelbl, anchor=west] at ({\xc+0.32}, \addy) (lp2) {$\ln\bm{p}_2^{(n)}$};
\node[edgelbl, anchor=west] at ({\xe+0.32}, \addy) (lp3) {$\ln\bm{p}_3^{(n)}$};
\node[edgelbl, anchor=west] at ({\xg+0.32}, \addy) (lpL) {$\ln\bm{p}_{T_{\max}}^{(n)}$};

\draw[arrow] (lp2.west) -- (add2.east);
\draw[arrow] (lp3.west) -- (add3.east);
\draw[arrow] (lpL.west) -- (addL.east);

% === Bus to CE loss ===
\pgfmathsetmacro{\cex}{\xh+0.3}

\draw[bus] (add2.south) -- (\xc, \busy);
\draw[bus] (add3.south) -- (\xe, \busy);
\draw[bus] (addL.south) -- (\xg, \busy);
\draw[bus] (\xc, \busy) -- (\xg, \busy);

\node[junction] at (\xe, \busy) {};
\node[junction] at (\xg, \busy) {};

\draw[arrow] (\xg, \busy) -- (\cex, \busy);

% === CE loss node ===
\node[block, fill=gray!5, anchor=west] at (\cex, \busy) (celoss) {Avg.\ CE};

\end{tikzpicture}
        \label{fig:pipeline_train}
    }
    \subfigure[Inference with parallel data queries.]{%
        % fig_pipeline_infer.tex
% Requires tikz libraries: arrows.meta, calc
\begin{tikzpicture}[
    >={Latex[length=1.6mm, width=1.4mm]},
    font=\footnotesize,
    line width=0.5pt,
    block/.style={draw, align=center, fill=white, inner sep=4pt,
                  rounded corners=2pt},
    data/.style={draw, thick, align=center, fill=white, inner sep=2pt,
                 rounded corners=2pt, minimum height=0.55cm, minimum width=0.5cm},
    query/.style={data, fill=mygreen!15, font=\scriptsize},
    label/.style={data, fill=myyellow!20, font=\scriptsize},
    addnode/.style={draw, thick, circle, minimum size=0.32cm,
                    inner sep=0pt, font=\scriptsize, fill=white},
    subnode/.style={draw, thick, circle, minimum size=0.32cm,
                    inner sep=0pt, font=\scriptsize, fill=white},
    arrow/.style={->, line width=0.5pt},
    bus/.style={line width=0.5pt},
    attn/.style={line width=0.3pt, black!48},
    edgelbl/.style={inner sep=1.5pt, font=\scriptsize},
    junction/.style={circle, fill=black, inner sep=0pt, minimum size=2.2pt}
]

% === Layout parameters ===
\def\ax{-3.6}
\def\dxp{0.65}
\def\dxq{0.95}
\def\ty{0}
\def\tboxtop{-0.55}
\def\tboxbot{-1.3}
\def\ley{-1.85}
\def\addy{-2.55}
\def\busy{-3.30}
\def\tfpad{0.6}

% Pilots
\pgfmathsetmacro{\xa}{\ax}
\pgfmathsetmacro{\xb}{\ax+\dxp}
\pgfmathsetmacro{\xdotsp}{\ax+2*\dxp}
\pgfmathsetmacro{\xc}{\ax+3*\dxp}
\pgfmathsetmacro{\xd}{\ax+4*\dxp}

% Separator
\pgfmathsetmacro{\xsep}{\xd+0.55}

% Queries
\pgfmathsetmacro{\xq}{\xsep+0.55}
\pgfmathsetmacro{\xr}{\xq+\dxq+0.5}
\pgfmathsetmacro{\xdotsq}{\xr+\dxq}
\pgfmathsetmacro{\xs}{\xdotsq+\dxq}

% === Pilot tokens ===
\node[query] at (\xa, \ty) {$\bm{z}_1$};
\node[label] at (\xb, \ty) {$\bm{\phi}_1$};
\node[edgelbl] at (\xdotsp, \ty) {$\cdots$};
\node[query] at (\xc, \ty) {$\bm{z}_{T_P}$};
\node[label] at (\xd, \ty) {$\bm{\phi}_{T_P}$};

% === Separator ===
\node[edgelbl] at (\xsep, \ty) {$\big|$};

% === Query tokens ===
\node[query] at (\xq, \ty) {$\bm{y}_1$\\$\bm{p}_1$};
\node[query] at (\xr, \ty) {$\bm{y}_2$\\$\bm{p}_2$};
\node[edgelbl] at (\xdotsq, \ty) {$\cdots$};
\node[query] at (\xs, \ty) {$\bm{y}_{T}$\\$\bm{p}_{T}$};

% === Transformer box ===
\pgfmathsetmacro{\tfl}{\ax-\tfpad}
\pgfmathsetmacro{\tfr}{\xs+\tfpad}
\fill[myred!8, rounded corners=2pt] (\tfl, \tboxtop) rectangle (\tfr-0.2, \tboxbot);
\draw[rounded corners=2pt] (\tfl, \tboxtop) rectangle (\tfr-0.2, \tboxbot);

% === Input arrows into transformer ===
\foreach \x in {\xa,\xb,\xc,\xd} {
    \draw[arrow] (\x, \ty-0.30) -- (\x, \tboxtop);
}
\foreach \x in {\xq,\xr,\xs} {
    \draw[arrow] (\x, \ty-0.30) -- (\x, \tboxtop);
}

% === Attention lines: every query token attends to all pilots and itself ===
\foreach \qx in {\xq, \xr, \xs} {
    \foreach \px in {\xa, \xb, \xc, \xd} {
        \draw[attn] (\px, \tboxtop) -- (\qx, \tboxbot);
    }
    \draw[attn] (\qx, \tboxtop) -- (\qx, \tboxbot);
}

% === Transformer label ===
\node[draw,anchor=west, font=\scriptsize, fill=white, inner sep=1.5pt,align=center]
    at ({\tfl-0.1}, \tboxbot) {Trained Transformer};

% === Residual logits (transformer output) ===
\node[edgelbl] at (\xq, \ley) {$\bm{l}_{\mathrm{e},1}^{(n)}$};
\node[edgelbl] at (\xr, \ley) {$\bm{l}_{\mathrm{e},2}^{(n)}$};
\node[edgelbl] at (\xdotsq, \ley) {$\cdots$};
\node[edgelbl] at (\xs, \ley) {$\bm{l}_{\mathrm{e},T}^{(n)}$};

\foreach \x in {\xq,\xr,\xs} {
    \draw[arrow] (\x, \tboxbot) -- (\x, \ley+0.18);
}

% === Adders: log-domain prior skip, posterior logits ===
\node[addnode] at (\xq, \addy) (add1) {$+$};
\node[addnode] at (\xr, \addy) (add2) {$+$};
\node[addnode] at (\xs, \addy) (addNs) {$+$};

\foreach \x/\add in {\xq/add1, \xr/add2, \xs/addNs} {
    \draw[arrow] (\x, \ley-0.18) -- (\add);
}

% === Log-prior labels ===
\node[edgelbl, anchor=east] at ({\xq-0.32}, \addy) (lp1) {$\ln\bm{p}_1^{(n)}$};
\node[edgelbl, anchor=east] at ({\xr-0.32}, \addy) (lp2) {$\ln\bm{p}_2^{(n)}$};
\node[edgelbl, anchor=east] at ({\xs-0.32}, \addy) (lpNs) {$\ln\bm{p}_{T}^{(n)}$};

\draw[arrow] (lp1.east) -- (add1.west);
\draw[arrow] (lp2.east) -- (add2.west);
\draw[arrow] (lpNs.east) -- (addNs.west);

% === Bus collecting posterior logits l_i^(n) ===
\draw[bus] (add1.south) -- (\xq, \busy);
\draw[bus] (add2.south) -- (\xr, \busy);
\draw[bus] (addNs.south) -- (\xs, \busy);
\draw[bus] (\xq, \busy) -- (\xs, \busy);

\node[junction] at (\xr, \busy) {};
\node[junction] at (\xs, \busy) {};

% === Symbol-to-bit conversion of the posterior estimate, Eq. (eq_probs_to_llr) ===
\node[anchor=east, draw, align=center, fill=white,inner sep=0.05cm,font=\scriptsize] (convert) at ({\xq-0.3}, \busy) {$L(c_k|\hat{\bm{q}}_i^{(n)}) \leftarrow \bm{l}_{i}^{(n)}$};

% === Left-side row: subtraction and decoder ===
\node[subnode, anchor=east] (sub) at ({\xq-2.9}, \busy) {$-$};

\node[block, fill=gray!5, left=1.0cm of sub] (decnode) {$\Pi^{-1}$\\Dec};

\draw[arrow] (\xq, \busy) -- (convert.east);
\draw[arrow] (convert.west) -- (sub.east);

% === Decoder-extrinsic LLR input ===
\draw[arrow] ($(sub.north)+(0,+0.4)$) node[edgelbl, above] {$L_{\mathrm{e}}^{\mathrm{dec}}(c_k)$}
    -- (sub.north);

% === Subtraction -> decoder ===
\draw[arrow] (sub.west) -- node[edgelbl, above] {$L_{\mathrm{e}}(c_k)$} (decnode.east);

\end{tikzpicture}
        \label{fig:pipeline_infer}
    }
    \caption{Training and inference pipelines. Gray lines indicate the attention connections permitted by the corresponding attention mask.}
    \label{fig:train_vs_inference}
\end{figure}

When a token vector $\bm{t}$ is assigned positional index $u$, its initial embedding is
\begin{equation}\label{eq_token_embedding}
    \bm{e}^{(0)} = \bm{W}_{\mathrm{emb}}\bm{t} + \bm{e}_{\mathrm{pos},u},
\end{equation}
where the learnable embedding matrix $\bm{W}_{\mathrm{emb}}\in\mathbb{R}^{D_E\times(2\max(N_r,N_t)+MN_t)}$ maps the token vector to a $D_E$-dimensional embedding, and $\bm{e}_{\mathrm{pos},u}\in\mathbb{R}^{D_E}$ is the learned embedding associated with positional index $u$~\cite{radford2019language}. The resulting embedding sequence is processed by $N_L$ Transformer layers, each with $N_H$ attention heads.

Let the final hidden representation $\bm{e}_{i}^{(N_L)}\in\mathbb{R}^{D_E}$ denote the output of the last Transformer layer corresponding to data query $i$. For stream $n$, the corresponding output head produces the residual logit vector
\begin{equation}\label{eq_extrinsic_logit}
    \bm{l}_{\mathrm{e},i}^{(n)} = \bm{W}_{O}^{(n)}\bm{e}_{i}^{(N_L)} + \bm{b}_{O}^{(n)},
\end{equation}
where the output matrix $\bm{W}_{O}^{(n)}\in\mathbb{R}^{M\times D_E}$ and the bias vector $\bm{b}_{O}^{(n)}\in\mathbb{R}^{M}$ are learnable. The posterior logit vector $\bm{l}_{i}^{(n)}$ and the posterior estimate $\hat{\bm{q}}_{i}^{(n)}$ are then obtained using the log-domain prior skip in~\eqref{eq_logit_decomp} and~\eqref{eq_residual_posterior}, respectively. The Transformer backbone, the stream-specific output heads, and the log-domain prior skip together constitute EXT-ICLNet.

\subsubsection{Training and Inference}\label{subsubse:train_infer}

As shown in Fig.~\ref{fig:pipeline_train}, offline training uses the training sequence $\mathcal{P}_{\mathrm{train}}$ defined in~\eqref{eq_prompt_train}, which contains $T_{\max}$ query-label pairs and hence $2T_{\max}$ tokens. The known label token $\bm{x}_{t}$ is placed after the corresponding query token $[\bm{y}_{t},\bm{p}_{t}]$, as in teacher forcing. The causal attention mask allows the representation at query position $t$ to depend on the first $t-1$ query-label pairs and the current query, but not on the current label or any subsequent tokens.

A single forward pass produces posterior estimates for all target positions $t\in\{2,\ldots,T_{\max}\}$. The first query-label pair is used as the initial context and is not included as a prediction target because no labeled context precedes its query. The empirical training loss is obtained by averaging the sum of the per-stream cross-entropies over the remaining target positions as
\begin{equation}\label{eq_training_loss}
    \hat{\mathcal{L}}_{\theta} = -\frac{1}{T_{\max}-1}\sum_{t=2}^{T_{\max}}\sum_{n=1}^{N_t}\ln\hat{q}_{t,\operatorname{idx}_{M}(x_t^{(n)})}^{(n)},
\end{equation}
where $\operatorname{idx}_{M}(x_t^{(n)})\in\{1,\ldots,M\}$ denotes the index of the transmitted symbol $x_t^{(n)}$ in the constellation $\mathcal{S}$. The empirical loss $\hat{\mathcal{L}}_{\theta}$ is the finite-sample counterpart of the population loss $\mathcal{L}_{\theta}$ in~\eqref{eq_expected_loss}. The detector parameters $\theta$ are optimized by minimizing the empirical loss.

At inference, the input sequence consists of the $2T_P$ tokens representing the pilot context $\mathcal{C}$, followed by the $T$ data-query tokens. No label tokens are included for the data queries. All data-query tokens are assigned the positional embedding $\bm{e}_{\mathrm{pos},2T_P+1}$ and attend to the complete pilot context, but not to one another. Consequently, processing the data queries jointly is equivalent to evaluating the per-symbol prompts $\mathcal{P}_i$ in~\eqref{eq_prompt_inference} separately. Increasing $T$ adds parallel query computations without introducing additional positional indices, so all $T$ data queries are processed in a single forward pass, as shown in Fig.~\ref{fig:pipeline_infer}. The detector-extrinsic LLRs then follow from $\hat{\bm{q}}_{i}^{(n)}$ in~\eqref{eq_residual_posterior} through~\eqref{eq_probs_to_llr} and~\eqref{eq_llr_chn_ext}.

\section{Numerical Evaluations}\label{se:results}

\subsection{Simulation Setup}\label{subse:setup}

We consider a $4\times4$ MIMO system ($N_t=N_r=4$) under i.i.d.\ Rayleigh block fading. Information bits are encoded using a 5G NR low-density parity-check (LDPC) code with rate matching~\cite{3gpp38212}, interleaved, modulated with the Gray-labeled 16-QAM of~\cite{3gpp38211} ($M=16$), and spatially multiplexed across the $N_t$ transmit streams following the index map $\pi(i,n,\kappa)$ in~\eqref{eq_llr_2_probs}. Unless otherwise stated, the codeword length is $N_c=1024$ and the code rate is $K/N_c=1/2$. No code-block segmentation or cyclic redundancy check attachment is applied, and the codeword is read out with redundancy version 0. The bit interleaver of~\cite{3gpp38212} is replaced by an independent random permutation for each codeword. The belief-propagation (BP) decoder uses flooding sum-product updates and runs up to 20 iterations per decoder call.
% \pp{PP: Provide mode details on the exact code and bit-to-symbol mapping used in order to make the results reproducible. Is the 3gpp reference sufficient for that or there are still degrees of freedom that need to be specified?}

EXT-ICLNet uses a GPT-2~\cite{radford2019language} Transformer backbone with $N_L=6$ layers, $N_H=8$ attention heads, embedding dimension $D_E=512$, and feed-forward hidden dimension 2048. Training uses $T_{\max}=80$ query-label pairs per prompt, corresponding to 160 tokens, and a fixed training set $\mathcal{H}_{\mathrm{train}}$ containing $N_{\mathrm{train}}=2^{15}$ channel realizations. The channel realizations are drawn independently from $\mathcal{D}_{\bm{H}}$, which generates i.i.d.\ complex Gaussian channel matrices with unit-variance entries. For each training prompt, the SNR in decibels and the quantization resolution are drawn independently and uniformly from $[0,30]$\,dB and $\{1,\ldots,32\}$, respectively, which defines $\mathcal{D}_{\sigma^2}^{\mathrm{train}}$ and $\mathcal{D}_B^{\mathrm{train}}$. The concentration-parameter distribution $\mathcal{D}_\alpha$ of Sec.~\ref{subse:extrinsic_sep} is $\log_{10}\alpha\sim\mathcal{U}[-2,0]$, which corresponds to $\alpha\in[0.01,1]$. Test data use a held-out channel set $\mathcal{H}_{\mathrm{test}}$ drawn independently from $\mathcal{D}_{\bm{H}}$ and satisfying $\mathcal{H}_{\mathrm{test}}\cap\mathcal{H}_{\mathrm{train}}=\emptyset$.

\subsection{Benchmarks}\label{subse:benchmarks}

We compare EXT-ICLNet against the following detectors.
\begin{itemize}[leftmargin=*,itemsep=1pt,topsep=2pt]
\item \emph{(B1) BLMMSE-PIC}~\cite{studer2011asic,bussgang1952crosscorrelation} is a model-based linear detector with parallel interference cancellation that accounts for quantization through a Bussgang-linearized model.
\item \emph{(B2) P-BEP}~\cite{santos2018turbo} is a model-based EP detector with $S_{\mathrm{EP}}=3$ self-iterations per turbo iteration. It assumes the unquantized linear-Gaussian model and does not account for quantization.
\item \emph{(B3) STS-SD}~\cite{studer2010soft} is a model-based soft-input soft-output sphere decoder, used as a near-maximum-likelihood (ML) reference.
\item \emph{(B4) OAMP-Net2}~\cite{he2020model} is a neural detector that unfolds $N_{\text{iter}}=6$ OAMP iterations. We extend it to accept the decoder priors in its MMSE denoiser and train it on i.i.d.\ Rayleigh channels with the same prior distribution as EXT-ICLNet.
\item \emph{(B5) EXT-GEPNet}~\cite{zhou2023graph} is a neural detector that enhances EP with a GNN and is trained with a three-step extrinsic procedure.
\item The MAP detectors with perfect CSI (\emph{MAP}) and with the pilots only (\emph{pilot-aided MAP})~\cite{taricco2005space}, whose complexity grows exponentially in $N_t$, are included as references. Both know $\sigma^2$. MAP also knows $\bm{H}$, whereas pilot-aided MAP knows only the channel distribution.
\end{itemize}
 
Each baseline is evaluated under two CSI regimes. \emph{Perfect CSI} assumes $\bm{H}$ and $\sigma^2$ are known exactly, which isolates the detector's intrinsic performance from estimation error. \emph{Estimated CSI} follows Sec.~\ref{subsubse:csi}, with an initial LS estimate from the $T_P$ pilots and decision-directed RLS refinement across turbo iterations~\cite{da2021adaptive,shiao2008combined}. The normalized mean square error of LS estimation is within $0.2$\,dB of that of MMSE estimation for $T_P=32$ and i.i.d.\ unit-variance channels.

\subsection{Performance Metrics}\label{subse:metrics}

We evaluate the detectors using the following metrics.

\emph{Post-decoding BER.} The BER is measured at the output of the channel decoder after a given number of turbo iterations. One-shot detection refers to the first iteration, in which the detector receives no a priori information from the decoder.

\emph{Turbo gain.} The improvement obtained through the turbo iterations is measured by the turbo gain $10\log_{10}(\mathrm{BER}_1/\mathrm{BER}_5)$, where $\mathrm{BER}_1$ and $\mathrm{BER}_5$ denote the post-decoding BERs after the first and fifth turbo iterations, respectively.

\emph{EXIT chart.} The EXIT function of a soft-input soft-output module gives the mutual information $I_E$ between the coded bits and the extrinsic LLRs at its output as a function of the mutual information $I_A$ between the coded bits and the a priori LLRs at its input~\cite{ten2001convergence}. The EXIT chart shows the EXIT function of the detector together with that of the decoder, where the latter is plotted with swapped axes because the extrinsic output of each module is the a priori input of the other. The exchange of information across turbo iterations follows a trajectory between the two curves, and the iterations improve the decoder output as long as the detector curve lies above the decoder curve. The EXIT chart therefore characterizes the quality of the extrinsic information produced by a detector, and it shows whether the detector can exploit increasingly reliable a priori information from the decoder.

\subsection{Robustness to Quantization and Pilot Length}\label{ssubse:quantization}

Both experiments fix SNR\,$=$\,10\,dB and 16-QAM, and report post-decoding BER under one-shot detection (no turbo iterations).

\begin{figure}[t]
    \centering
    \includegraphics[width=\columnwidth]{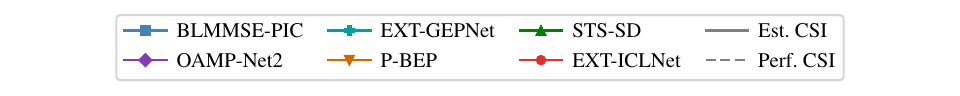}\\
    \vspace{-2mm}
    \subfigure[]{%
        \includegraphics[width=0.48\columnwidth]{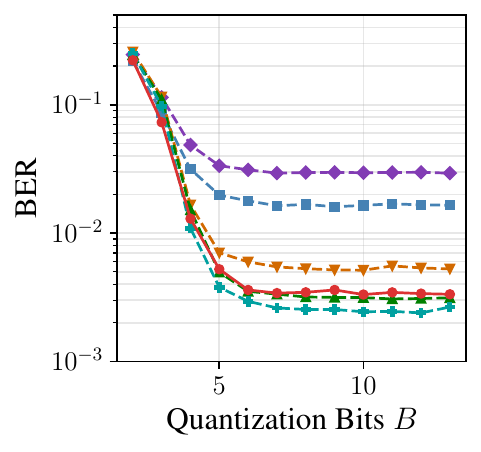}
        \label{fig:ber_bits}
    }%
    \hfill%
    \subfigure[]{%
        \includegraphics[width=0.48\columnwidth]{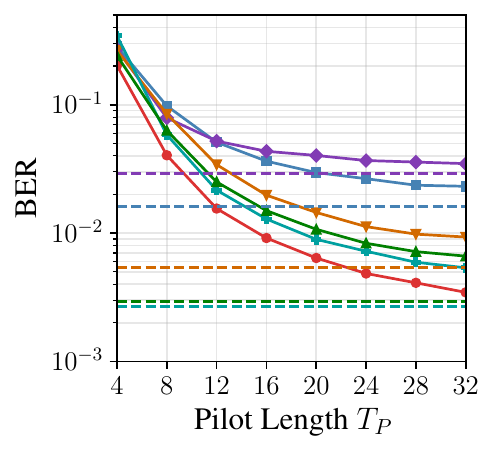}
        \label{fig:ber_pilot}
    }
    \caption{Post-decoding BER at SNR\,$=$\,10\,dB under 16-QAM, one-shot detection: (a)~impact of quantization resolution $B$ at $T_P = 32$ with all baselines using perfect CSI; (b)~impact of pilot length $T_P$ at $B = 8$, with baselines using LS-estimated CSI (solid) and perfect CSI (dashed).}
    \label{fig:ber_bits_pilot}
\end{figure}

Fig.~\ref{fig:ber_bits} shows the impact of the quantization resolution $B$, with all baselines using perfect CSI. At $B=3$, EXT-ICLNet achieves the lowest BER among all detectors, although it uses no channel estimate and the baselines use perfect CSI. The linear-Gaussian and Bussgang-linearized models of the baselines are least accurate at coarse quantization, whereas EXT-ICLNet learns the quantized observation model from training prompts that span the range of quantization resolutions. The BER of all detectors saturates for $B\gtrsim6$, where EXT-ICLNet matches perfect-CSI STS-SD and only perfect-CSI EXT-GEPNet attains a lower BER. We use $B=8$ in the remaining experiments.

Fig.~\ref{fig:ber_pilot} shows the impact of the pilot length $T_P$ at $B=8$, with the estimated-CSI and perfect-CSI baselines represented by solid and dashed curves, respectively. EXT-ICLNet achieves a lower BER than all estimated-CSI baselines at every pilot length, and its BER decreases more rapidly with $T_P$. The estimated-CSI baselines plug in a channel estimate whose error they do not account for, whereas EXT-ICLNet targets the posterior~\eqref{eq_true_posterior}, which accounts for the channel uncertainty that remains after the pilots are observed. With $T_P=12$ pilots, EXT-ICLNet matches perfect-CSI BLMMSE-PIC, and with $T_P=24$ it outperforms perfect-CSI P-BEP. We use $T_P=32$ in the remaining experiments.

\subsection{Post-Decoding BER vs.\ SNR}\label{ssubse:snr}
\begin{figure}[t]
    \centering
    \includegraphics[width=\columnwidth]{figures/t1_ber_snr/legend.pdf}\\
    \vspace{-2mm}
    \subfigure[Baselines: Perf.\ CSI, iter.\ 1]{%
        \includegraphics[width=0.48\columnwidth]{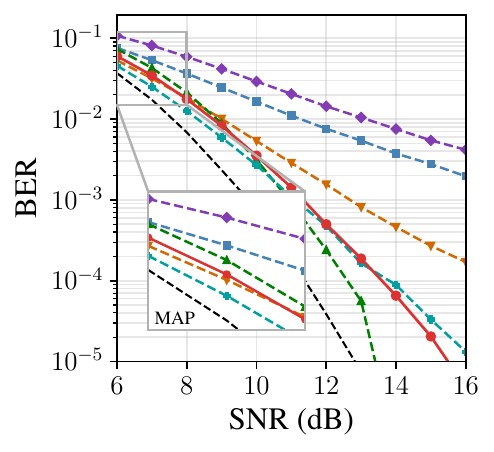}
        \label{fig:ber_snr_perf_i1}
    }%
    \hfill%
    \subfigure[Baselines: Perf.\ CSI, iter.\ 5]{%
        \includegraphics[width=0.48\columnwidth]{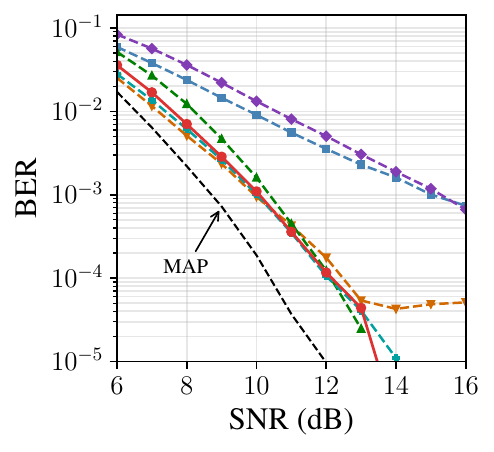}
        \label{fig:ber_snr_perf_i5}
    }\\
    \subfigure[Baselines: Est.\ CSI, iter.\ 1]{%
        \includegraphics[width=0.48\columnwidth]{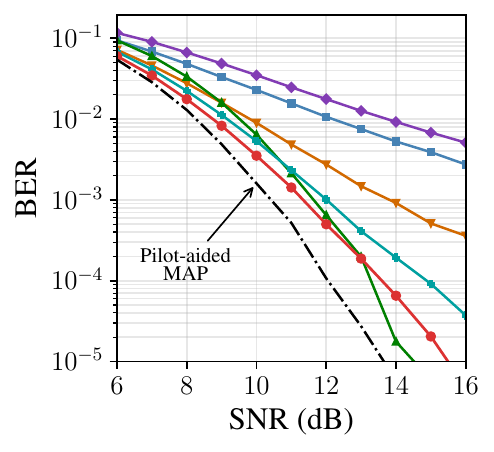}
        \label{fig:ber_snr_est_i1}
    }%
    \hfill%
    \subfigure[Baselines: Est.\ CSI, iter.\ 5]{%
        \includegraphics[width=0.48\columnwidth]{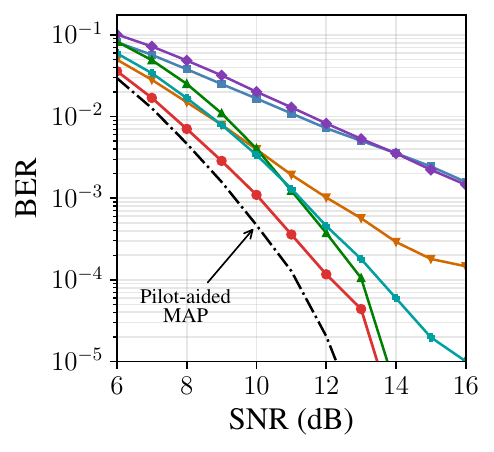}
        \label{fig:ber_snr_est_i5}
    }
\caption{Post-decoding BER vs.\ SNR under 16-QAM, rate~$1/2$, $B{=}8$, $T_P{=}32$. Top row: baselines with perfect CSI; bottom row: baselines with estimated CSI. Black dashed and dash-dotted curves show MAP and pilot-aided MAP, respectively.}
    \label{fig:ber_snr}
\end{figure}
 
Figs.~\ref{fig:ber_snr_perf_i1} and~\ref{fig:ber_snr_perf_i5} compare EXT-ICLNet with the perfect-CSI baselines. At the first turbo iteration, EXT-ICLNet outperforms P-BEP above approximately $8$\,dB and EXT-GEPNet above approximately $14$\,dB, while STS-SD achieves a lower BER above approximately $10$\,dB. EXT-ICLNet achieves a positive turbo gain at every SNR, whereas P-BEP reaches an error floor of approximately $5\times10^{-5}$ above 13\,dB after the fifth iteration.

Figs.~\ref{fig:ber_snr_est_i1} and~\ref{fig:ber_snr_est_i5} present the corresponding comparison with estimated-CSI baselines. After the fifth iteration, EXT-ICLNet reaches a BER of $10^{-3}$ at approximately 10\,dB and approaches pilot-aided MAP within approximately 0.6\,dB, whereas estimated-CSI EXT-GEPNet requires approximately 11\,dB. Compared with estimated-CSI STS-SD, EXT-ICLNet achieves a lower BER up to approximately $13$\,dB after the first iteration, while the crossover moves above $14$\,dB after the fifth iteration. 

\subsection{Extrinsic Information Quality and Robustness}\label{subse:exit_robustness}

\begin{figure}[t]
    \centering
    \includegraphics[width=0.85\columnwidth]{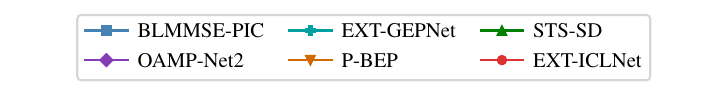}\\
    \vspace{-3mm}
    \subfigure[EXIT chart, SNR $=$ 4.5\,dB]{%
        \includegraphics[width=0.48\columnwidth]{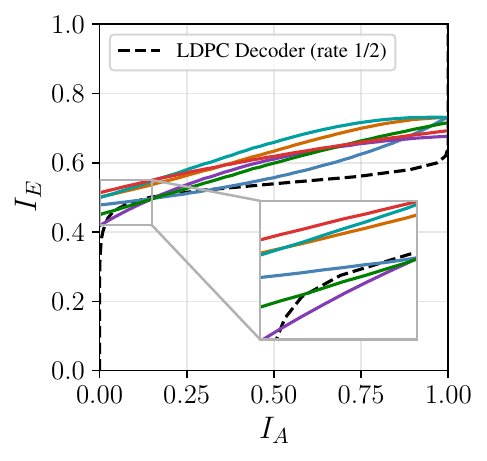}
        \label{fig:exit_4dB}
    }%
    \hfill%
    \subfigure[EXIT chart, SNR $=$ 10\,dB]{%
        \includegraphics[width=0.48\columnwidth]{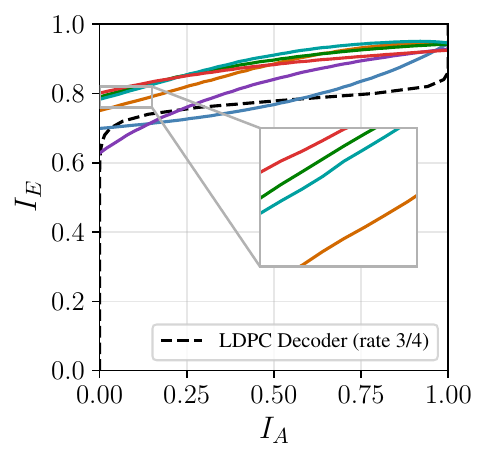}
        \label{fig:exit_10dB}
    }\\
    \vspace{-2mm}
    \subfigure[EXIT chart, SNR $=$ 10\,dB]{%
        \includegraphics[width=0.48\columnwidth]{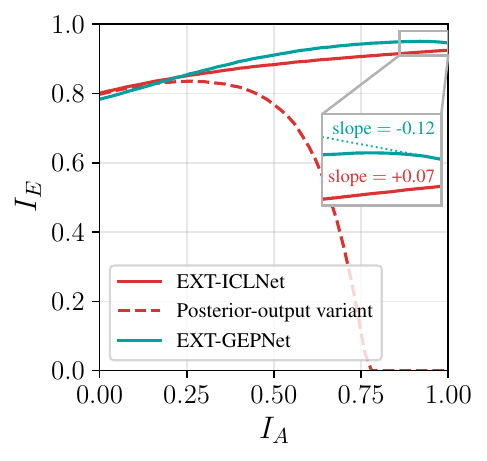}
        \label{fig:exit_arch}
    }%
    \hfill%
    \subfigure[BER vs.\ $\gamma$]{%
        \includegraphics[width=0.45\columnwidth]{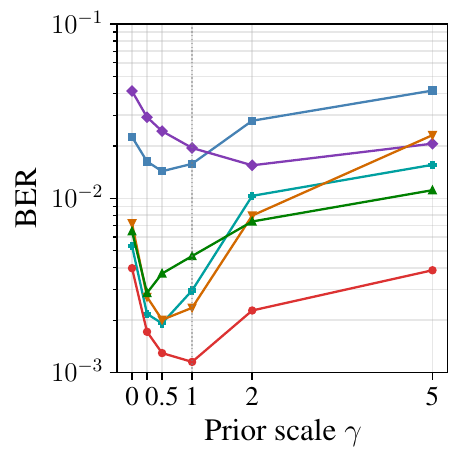}
        \label{fig:robust_post}
    }
    \caption{Extrinsic information quality at $B = 8$, $T_P = 32$, 16-QAM, $4 \times 4$. (a)--(b)~EXIT charts comparing EXT-ICLNet against baselines with LS-estimated CSI, with the rate-$1/2$ and rate-$3/4$ decoder characteristics, respectively; the LDPC decoder characteristic is plotted with swapped axes. (c)~EXIT curves at SNR\,$=$\,10\,dB; inset zooms the high-$I_A$ region. (d)~Post-decoding BER after the 5th turbo iteration vs.\ prior scale $\gamma$ at SNR\,$=$\,6\,dB.}
    \label{fig:exit_robustness}
\end{figure}

For each a priori mutual information $I_A$, the EXIT charts are obtained by generating synthetic prior LLRs from the consistent Gaussian model~\cite{ten2001convergence} and measuring the extrinsic mutual information $I_E$ at the detector output. The baselines use LS-estimated CSI.

Fig.~\ref{fig:exit_4dB} shows the EXIT chart at SNR\,$=$\,4.5\,dB with the rate-$1/2$ decoder characteristic. EXT-ICLNet and P-BEP leave the widest tunnels to the decoder characteristic among the evaluated detectors, and the same two detectors retain the largest turbo gains at rate~$2/3$ in Sec.~\ref{subse:coderate}. As $I_A$ increases, the transfer curve of P-BEP flattens, whereas that of EXT-ICLNet continues to increase.

Fig.~\ref{fig:exit_10dB} shows the EXIT chart at SNR\,$=$\,10\,dB with the rate-$3/4$ decoder characteristic. At $I_A=0$, EXT-ICLNet attains the highest $I_E$ among the evaluated detectors, and its EXIT curve increases over the evaluated range.

Fig.~\ref{fig:exit_arch} isolates the effect of the prior skip by comparing EXT-ICLNet with the posterior-output variant of Sec.~\ref{subse:residual_head} and with EXT-GEPNet. The EXIT curve of the posterior-output variant peaks near $I_A=0.25$ and falls to zero for $I_A\gtrsim0.78$. In this variant, the prior enters the posterior only through the learned logits, so the division $\hat{\ell}_{i}^{(n)}(s_m) \propto \hat{q}_{i,m}^{(n)}/p_{i,m}^{(n)}$ amplifies the approximation error wherever the prior probabilities are small, and small prior probabilities become common at high $I_A$. In EXT-ICLNet, the own-stream prior enters the posterior exactly through~\eqref{eq_logit_decomp}, and the subtraction in~\eqref{eq_llr_chn_ext} removes the a priori LLR of each bit without dividing by a learned approximation of the prior, so its EXIT curve increases monotonically over the evaluated range. EXT-GEPNet attains a higher $I_E$ than EXT-ICLNet for $I_A\gtrsim0.25$ but exhibits a negative slope over $I_A\in[0.98,1.0]$, whereas EXT-ICLNet retains a positive slope in this interval.

Fig.~\ref{fig:robust_post} scales the decoder-extrinsic LLRs by a factor $\gamma$ before their conversion to symbol priors in~\eqref{eq_llr_2_probs}, so that $\gamma>1$ yields overconfident priors and $\gamma<1$ yields underconfident priors. EXT-ICLNet achieves the lowest BER at every $\gamma$ and attains its minimum at $\gamma=1$. Because each training symbol is drawn from its prior PMF (Sec.~\ref{subse:extrinsic_sep}), EXT-ICLNet is trained with calibrated priors, which the prior skip in~\eqref{eq_logit_decomp} adds to the residual logits with unit weight, so the unscaled decoder LLRs at $\gamma=1$ match the training condition. In contrast, BLMMSE-PIC, P-BEP, and EXT-GEPNet attain their minima at $\gamma=0.5$ and STS-SD at $\gamma=0.25$, which indicates that these detectors overweight the decoder feedback, whereas OAMP-Net2 attains its minimum at $\gamma=2$ and therefore underweights it. For $\gamma>1$, the BER of P-BEP and EXT-GEPNet rises to more than eight times its minimum at $\gamma=5$, whereas that of EXT-ICLNet rises by a factor of about three.

\subsection{Impact of Channel Code Rate}\label{subse:coderate}

We evaluate the effect of the LDPC code rate $K/N_c$ at two SNR levels. All baselines use estimated CSI. We report the turbo gain defined in Sec.~\ref{subse:metrics}.
 
\begin{figure}[t]
    \centering
    \includegraphics[width=0.85\columnwidth]{figures/t6_ber_coderate/legend.pdf}\\
    \vspace{-3mm}
    \subfigure[Turbo gain, SNR\,$=$\,4.5\,dB]{%
        \includegraphics[width=0.48\columnwidth]{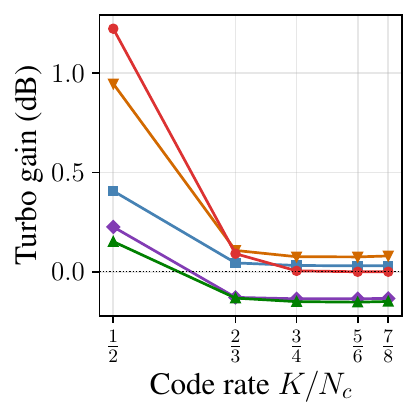}
        \label{fig:turbo_gain_5}
    }
    \hfill
    \subfigure[Turbo gain, SNR\,$=$\,10\,dB]{%
        \includegraphics[width=0.48\columnwidth]{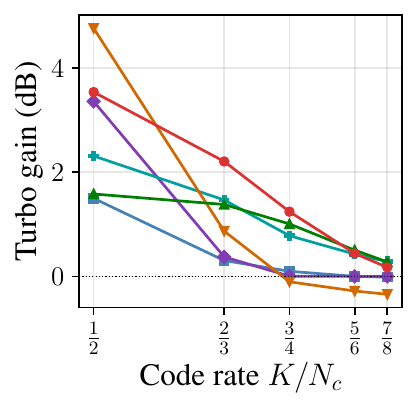}
        \label{fig:turbo_gain_10}
    }
    \caption{Turbo gain $10\log_{10}(\mathrm{BER}_1/\mathrm{BER}_5)$ versus LDPC code rate (16-QAM, $4\times4$, $B=8$, $T_P=32$, $N_c=1024$). All baselines use estimated CSI.}
    \label{fig:coderate}
\end{figure}

Fig.~\ref{fig:turbo_gain_5} shows the turbo gain at SNR\,$=$\,4.5\,dB. At code rate~$1/2$, EXT-ICLNet achieves the largest gain of 1.22\,dB, followed by P-BEP. The gains of all detectors decrease as the code rate increases. At rate~$2/3$, EXT-ICLNet and P-BEP retain the largest positive gains, consistent with their EXIT curves in Fig.~\ref{fig:exit_4dB}. At higher code rates, the gain of EXT-ICLNet remains close to $0$\,dB, while STS-SD and OAMP-Net2 exhibit negative gains.

Fig.~\ref{fig:turbo_gain_10} shows the turbo gain at SNR\,$=$\,10\,dB. EXT-ICLNet maintains a positive gain at every code rate and achieves the largest gain at rates~$2/3$ and~$3/4$. P-BEP achieves the largest gain at rate~$1/2$, but its gain becomes negative at rates~$\geq3/4$.

\subsection{Generalization to Spatially Correlated Channels}\label{subse:correlated}
 
The experiments thus far use i.i.d.\ Rayleigh fading. We next evaluate all detectors under the Kronecker correlation model $\bm{H} = \bm{R}_r^{1/2} \bm{U} \bm{R}_t^{1/2}$, where $\bm{U}$ is an i.i.d.\ Rayleigh matrix and $\bm{R}_r$, $\bm{R}_t$ are exponential correlation matrices with coefficient $\rho$. We set $\rho = 0.5$ to match the configuration of~\cite{zhou2023graph}. The EXT-ICLNet model trained on i.i.d.\ Rayleigh channels is applied directly to the correlated channels without retraining.
  
\begin{figure}[t]
    \centering
    \includegraphics[width=0.85\columnwidth]{figures/t6_ber_coderate/legend.pdf}\\
    \vspace{-3mm}
    \subfigure[Iter.\ 1]{%
        \includegraphics[width=0.48\columnwidth]{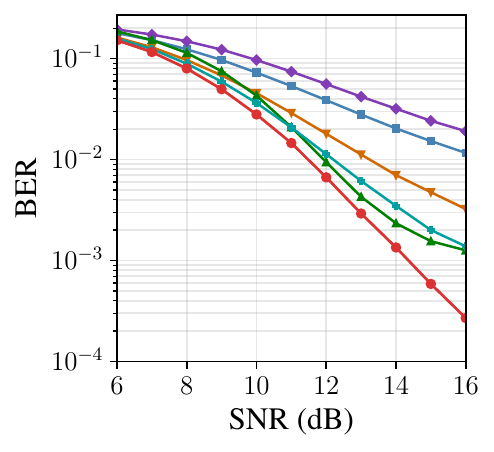}
        \label{fig:ber_corr_i1}
    }%
    \hfill%
    \subfigure[Iter.\ 5]{%
        \includegraphics[width=0.48\columnwidth]{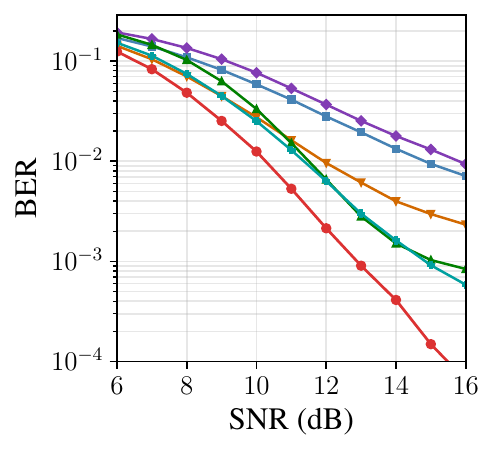}
        \label{fig:ber_corr_i5}
    }
    \caption{Post-decoding BER versus SNR under the Kronecker correlated channel model with $\rho=0.5$ (16-QAM, $4\times4$, $B=8$, $T_P=32$, rate $1/2$). Baselines use estimated CSI.}
    \label{fig:ber_correlated}
\end{figure}
 
In Fig.~\ref{fig:ber_correlated}, EXT-ICLNet achieves the lowest BER among all evaluated detectors, and its advantage grows with SNR. At 14\,dB after the fifth iteration, its BER is about four times lower than those of EXT-GEPNet and STS-SD, the best baselines, and more than one order of magnitude lower than those of BLMMSE-PIC and OAMP-Net2. The pilot context conveys the correlated channel to the detector at inference.

\subsection{Impact of Training-Channel Diversity}\label{ssubse:generalization}

\begin{figure}[t]
    \centering
    \subfigure[Coded BER]{%
        \includegraphics[width=0.49\columnwidth]{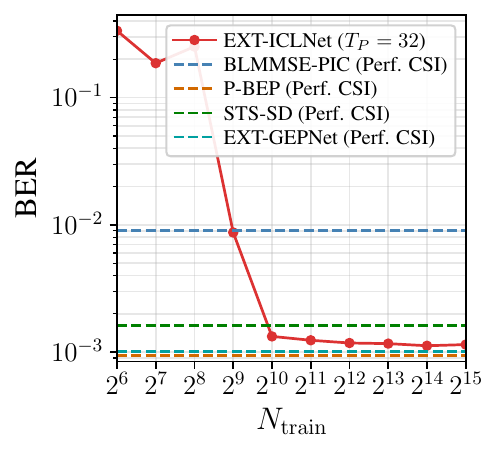}
        \label{fig:ber_ntask_ber}
    }%
    \hfill%
    \subfigure[EXIT mutual information]{%
        \includegraphics[width=0.47\columnwidth]{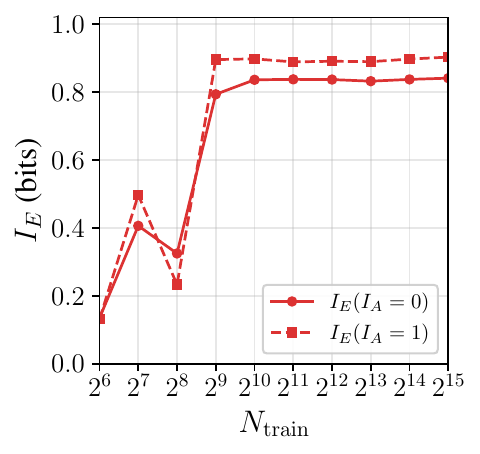}
        \label{fig:ber_ntask_ie}
    }
    \caption{EXT-ICLNet versus the number of distinct training channel realizations $N_{\text{train}}$ at SNR\,$=$\,10\,dB (16-QAM, $4\times4$, $B=8$, $T_P=32$, rate-$1/2$ LDPC). (a)~Post-decoding BER after the fifth turbo iteration, together with perfect-CSI baselines. (b)~EXIT mutual information at the uniform ($I_A=0$) and one-hot ($I_A=1$) prior endpoints on a held-out probe set.}
    \label{fig:ber_ntask}
\end{figure}

We vary $N_{\text{train}}$, the size of the fixed training channel set $\mathcal{H}_{\mathrm{train}}$, from $2^6$ to $2^{15}$ and retrain EXT-ICLNet from scratch for each value. Each model is evaluated at SNR\,$=$\,10\,dB using the held-out channel set $\mathcal{H}_{\mathrm{test}}$. Fig.~\ref{fig:ber_ntask_ber} shows the effect of the number of distinct training channel realizations on the BER. The BER drops by more than an order of magnitude as $N_{\text{train}}$ increases from $2^{8}$ to $2^{10}$ and changes little beyond $2^{11}$. For $N_{\text{train}}\geq2^{10}$, the BER of EXT-ICLNet falls below that of perfect-CSI STS-SD and approaches those of perfect-CSI P-BEP and EXT-GEPNet. Below $N_{\text{train}}=2^{10}$, the model does not generalize to the held-out channels; above it, the number of distinct training channels no longer limits performance.

Fig.~\ref{fig:ber_ntask_ie} shows the corresponding EXIT mutual information at $I_A=0$ and $I_A=1$. Both values increase as $N_{\text{train}}$ increases from $2^{8}$ to $2^{10}$ and change little for $N_{\text{train}}\geq2^{10}$. In this range, $I_E(I_A=1)$ exceeds $I_E(I_A=0)$, so EXT-ICLNet extracts more extrinsic information when the a priori information on the other bits is perfect than when it is absent, because perfect a priori information removes the interference from the other streams.

\subsection{Computational Complexity}\label{subse:complexity}

Table~\ref{tab:complexity} lists the complexity of each detector per turbo iteration. For the estimated-CSI baselines, channel estimation adds $\mathcal{O}\big((T_P+T)N^3\big)$ per iteration. In Table~\ref{tab:complexity}, $\bar{\nu}$ denotes the average number of nodes visited per symbol by STS-SD, and $T_G$, $L_G$, and $N_h$ denote the number of EP iterations, the number of GNN message-passing rounds, and the hidden-layer width of the GNN in EXT-GEPNet, respectively. EXT-ICLNet has no separate channel-estimation stage and processes the $2T_P$ pilot tokens and the $T$ data-query tokens in a single forward pass. Its complexity is dominated by the per-token projections and feed-forward layers, of order $N_LD_E^2$, while $N$ and $M$ enter only through the embedding and output layers. For fixed model dimensions and pilot length, the complexity of EXT-ICLNet therefore grows linearly in $N$ and $M$, whereas those of MAP and pilot-aided MAP grow exponentially in $N$ and those of the EP-based and linear detectors grow polynomially in $N$. The pilot tokens do not attend to the data queries, so their representations can be computed once per coherence block and reused across turbo iterations. Within each forward pass, all data queries are processed in parallel through dense matrix products.

\begin{table}[t]
\footnotesize
\centering
\caption{Computational complexity per turbo iteration, with $N\coloneqq\max(N_t,N_r)$.}
\label{tab:complexity}
\setlength{\tabcolsep}{4pt}
\renewcommand{\arraystretch}{1.15}
\begin{tabular}{|l|l|}
\hline
\textbf{Detector} & \textbf{Complexity} \\
\hline
BLMMSE-PIC & $\mathcal{O}\big(T(N^4+NM)\big)$ \\
P-BEP & $\mathcal{O}\big(S_{\mathrm{EP}}T(N^3+NM)\big)$ \\
OAMP-Net2 & $\mathcal{O}\big(N_{\mathrm{iter}}T(N^3+NM)\big)$ \\
EXT-GEPNet & $\mathcal{O}\big(T_GT(N^3+L_GN^2N_h^2+NM)\big)$ \\
STS-SD & $\mathcal{O}\big(T(\bar{\nu}N+NM)\big)$, $\bar{\nu}\le M^{N}$ \\
MAP & $\mathcal{O}\big(TM^{N}N\big)$ \\
Pilot-aided MAP & $\mathcal{O}\big((T+N^2)M^{N}N\big)$ \\
\hline
EXT-ICLNet & $\mathcal{O}\big((T_P+T)(N_LD_E^2+N_LT_PD_E+NMD_E)\big)$ \\
\hline
\end{tabular}
\end{table}

\section{Conclusion}\label{se:conclusion}

We proposed EXT-ICLNet, an ICL-based soft-input soft-output detector for MIMO turbo receivers. EXT-ICLNet takes the pilots directly as in-context examples and is trained to approximate the symbol posterior given the pilots, a target that accounts for the channel uncertainty that a point channel estimate ignores. Because the network also takes the a priori information as input, a log-domain prior skip adds the a priori log-probabilities to the network output, and we showed that the network output then targets a likelihood term that does not depend on the a priori information of its own stream. The extrinsic LLRs therefore follow from single-stage posterior training.

Numerical results for a $4\times4$ MIMO system with 16-QAM and a quantized front end show that EXT-ICLNet achieves a lower BER than the baselines with estimated CSI at low-to-medium SNRs and than the EP-based baselines with perfect CSI at high SNRs. Its EXIT curve increases monotonically, whereas those of EXT-GEPNet and of the posterior-output variant without the prior skip decrease at high a priori mutual information. A model trained on i.i.d.\ Rayleigh channels achieves the lowest BER under spatially correlated channels without retraining, and for fixed model dimensions and pilot length, the complexity of EXT-ICLNet grows linearly in the numbers of antennas and constellation points.
 
\bibliography{reference}
\end{document}